\documentclass[reprint,amssymb,amsmath,aps,superscriptaddress,floatfix,pra]{revtex4-1}

\pdfoutput=1
\usepackage{graphicx}
\usepackage{amssymb}
\usepackage{epstopdf}
\usepackage{upgreek}
\usepackage{dcolumn}
\usepackage{bm}
\usepackage{gensymb}
\usepackage{braket}
\usepackage{amsmath}
\usepackage{float}
\usepackage{sidecap}
\usepackage{graphicx}
  
\expandafter\ifx\csname package@font\endcsname\relax\else
 \expandafter\expandafter
 \expandafter\usepackage
 \expandafter\expandafter
 \expandafter{\csname package@font\endcsname}%
\fi

\newcommand{\micron}{~$\upmu$m\ }

\newcommand{\be}{\begin{eqnarray}}
\newcommand{\ee}{\end{eqnarray}}
\newcommand{\bfig}{\begin{figure}}
\newcommand{\efig}{\end{figure}}

\begin{document}

\title{Widely tunable on-chip microwave circulator for superconducting quantum circuits}
\author{Benjamin J. Chapman}
\email{benjamin.chapman@colorado.edu}
\affiliation{JILA, National Institute of Standards and Technology and the University of Colorado, Boulder, Colorado 80309, USA}
\affiliation{Department of Physics, University of Colorado, Boulder, Colorado 80309, USA}
\author{Eric I. Rosenthal}
\affiliation{JILA, National Institute of Standards and Technology and the University of Colorado, Boulder, Colorado 80309, USA}
\affiliation{Department of Physics, University of Colorado, Boulder, Colorado 80309, USA}
\author{Joseph Kerckhoff}
\altaffiliation{Current address: HRL Laboratories, LLC, Malibu, CA 90265, USA}
\affiliation{JILA, National Institute of Standards and Technology and the University of Colorado, Boulder, Colorado 80309, USA}
\affiliation{Department of Physics, University of Colorado, Boulder, Colorado 80309, USA}
\author{Bradley A. Moores}
\affiliation{JILA, National Institute of Standards and Technology and the University of Colorado, Boulder, Colorado 80309, USA}
\affiliation{Department of Physics, University of Colorado, Boulder, Colorado 80309, USA}
\author{Leila R. Vale}
\affiliation{National Institute of Standards and Technology, Boulder, Colorado 80305, USA}
\author{J.~A.~B. Mates}
\affiliation{National Institute of Standards and Technology, Boulder, Colorado 80305, USA}
\author{Gene C. Hilton}
\affiliation{National Institute of Standards and Technology, Boulder, Colorado 80305, USA}
\author{Kevin Lalumi\`ere}
\altaffiliation{Current address: Anyon Systems Inc., Dorval, Qu\'ebec H9P 1G9, Canada  }
\affiliation{D\'epartement de Physique, Universit\'e de Sherbrooke, Sherbrooke, Qu\'ebec J1K 2R1, Canada}
\author{Alexandre Blais}
\affiliation{D\'epartement de Physique, Universit\'e de Sherbrooke, Sherbrooke, Qu\'ebec J1K 2R1, Canada}
\affiliation{Canadian Institute for Advanced Research, Toronto, Ontario M5G 1Z8, Canada}
\author{K.~W. Lehnert}
\affiliation{JILA, National Institute of Standards and Technology and the University of Colorado, Boulder, Colorado 80309, USA}
\affiliation{Department of Physics, University of Colorado, Boulder, Colorado 80309, USA}
\date{\today}

\begin{abstract}
We report on the design and performance of an on-chip microwave circulator with a widely (GHz) tunable operation frequency.  Non-reciprocity is created with a combination of frequency conversion and delay, and requires neither permanent magnets nor microwave bias tones, allowing on-chip integration with other superconducting circuits without the need for high-bandwidth control lines. Isolation in the device exceeds 20 dB over a bandwidth of tens of MHz, and its insertion loss is small, reaching as low as 0.9 dB at select operation frequencies.  Furthermore, the device is linear with respect to input power for signal powers up to hundreds of fW ($\approx 10^3$ circulating photons), and the direction of circulation can be dynamically reconfigured.  We demonstrate its operation at a selection of frequencies between 4 and 6 GHz.
\end{abstract}

\maketitle

\section{Introduction}
In recent years, experiments on one or several superconducting qubits have shown that the circuit quantum electrodynamics architecture~\cite{blais:2004} is a viable platform for the realization of a quantum information processor~\cite{kelly:2015,ofek:2016}.  This success is in part due to the advent of high-quality microwave amplifiers~\cite{castellanos:2008,bergeal:2010,macklin:2015}, which allow for near-quantum limited amplification and single-shot, quantum-non-demolition readout of quantum states~\cite{vijay:2011,riste:2012}.

\begin{figure}[!htb]
\begin{center}
\includegraphics[width=1\linewidth]{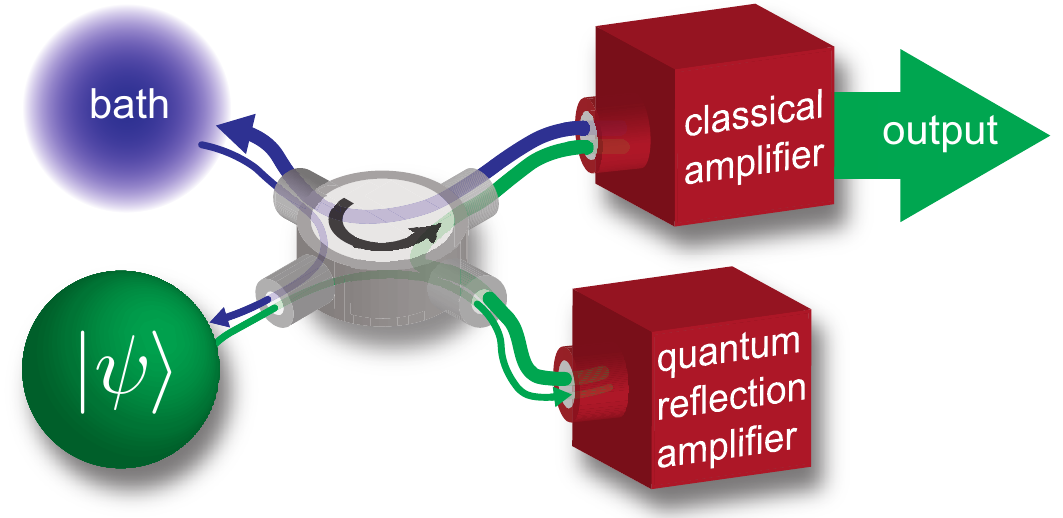}
\caption {A four-port circulator (gray cylinder) routes fields incident on one of its port out a neighboring port, as indicated by the arrow.   In circuit quantum electrodynamics, these devices are used to separate incoming and outgoing fields for quantum reflection amplifiers, and to isolate fragile quantum systems from the back-action of classical amplifiers.  Here we use ``quantum'' and ``classical'' to mean amplifiers with added noise equal to and much greater than half a photon, respectively~\cite{caves:1982}.
}
\label{fig:Motivation3D}
\end{center}
\end{figure}

As superconducting qubit experiments scale in complexity, further signal processing innovations are needed to preserve the high level of control demonstrated in few-qubit experiments. One bottle-neck in this area is the task of unidirectional signal routing.  Enforcing one-way signal propagation is critical, for example, in the separation of incoming and outgoing fields for quantum-limited reflection amplifiers, or the isolation of sensitive quantum devices from the back-action of the microwave receiver (Fig.~\ref{fig:Motivation3D}).


Currently, these tasks are performed by commercial ferrite junction circulators. These devices violate Lorentz reciprocity---the symmetry, in an electrical network, under exchange of source and detector~\cite{pozar:2011}---with large permanent magnets ($\approx 1$ mT stray fields) and the Faraday effect~\cite{fay:1965}.  Unfortunately, their size and reliance on these magnets make ferrite circulators difficult to integrate on-chip with superconducting circuits, and unattractive for long-term applications in networks with many superconducting qubits.  


Recognition of the need for a scalable circulator has therefore motivated the investigation of alternate means for generating non-reciprocity, using, for example, the quantum Hall effect~\cite{viola:2014,mahoney:2017,mahoney:2017b} and active devices~\cite{anderson:1965,anderson:1966,kamal:2011,metelmann:2015,kerckhoff:2015,abdo:2014,estep:2014,ranzani:2015,sliwa:2015,lecocq:2017,abdo:2017,bernier:2017,peterson:2017,barzanjeh:2017,metelmann:2017,khorasani:2017}. All of these approaches are chip-based, or can be adapted for chip-based implementations.  Scalability, however, requires more than miniaturization.  An ideal replacement technology would be both monolithic \emph{and} operable without high-bandwidth control lines, which are a limited resource in cryogenic microwave experiments.  It would also be low loss, linear at power levels typical for qubit readout and amplification, and flexible, in the sense that it should either be broadband, like a commercial circulator, or tunable over a wide frequency range, like some Josephson parametric amplifiers~\cite{castellanos:2007,mutus:2013}.


Here we present the performance of a superconducting microwave circulator proposed in Ref.~\cite{kerckhoff:2015}, which meets these stringent requirements.  The non-reciprocity is created with a combination of frequency conversion and delay, which ultimately preserves the frequency of the input signal.  Its operation requires no microwave frequency control tones, its center frequency may be tuned over a range of several GHz, and the device is realized on a 4 mm chip fabricated with a high-yield 
Nb/AlO$_x$/Nb trilayer process~\cite{sauvageau:1995,mates:2008}.  We first describe its theory of operation and its superconducting realization.  We then present experimental results, including measurements of the circulator's scattering matrix elements and a characterization of its transmission spectrum and linearity.  These measurements are performed over a range of different operation frequencies and with the circulator configured for clockwise and counterclockwise circulation, highlighting the device's tunability and the capability to dynamically reconfigure its sense of circulation \emph{in-situ}.

\section{Theory of Operation}
\label{sec:modelsystem}

\begin{figure*}[hbt]
\begin{center}
\includegraphics[width=1.0\linewidth]{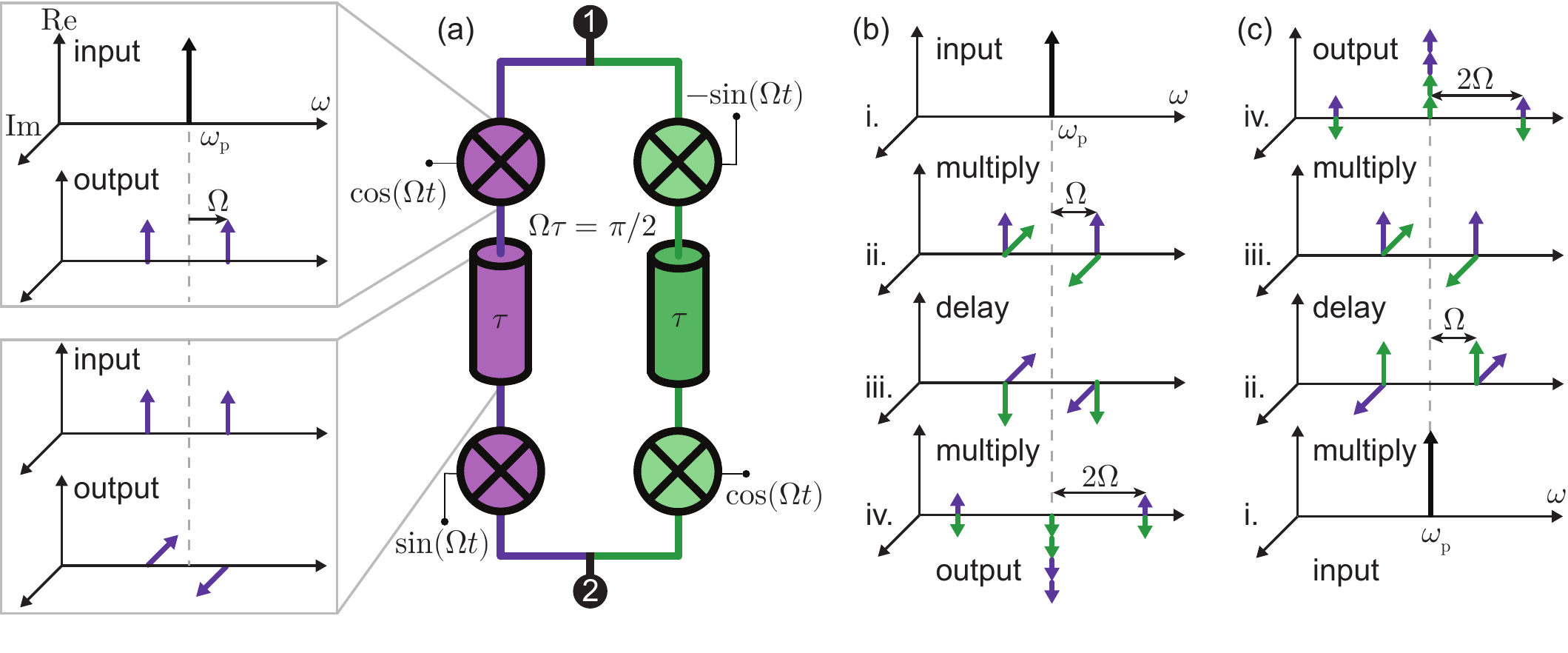}
\caption {Conceptual diagram of non-reciprocity generated with frequency conversion and delay.  (a) Lumped-element network that forms a gyrator. The insets show how fields are transformed by the network's two components: multiplying elements and delays.   In the upper inset, an input field at frequency $\omega_p$ (top panel) is multiplied by $\cos(\Omega t)$ to create a field with spectral components at $\omega_p \pm \Omega$ (bottom panel).  The real and imaginary axes of the plot shows the phase of these spectral components in a frame rotating at $\omega_p$.  In the lower inset, a delay of length $\tau = \pi/2\Omega$ advances or retards the phase of spectral components at $\omega_p + \Omega$ or $\omega_p - \Omega$ by $\pi/2$. (b) Calculation of the forward-scattering parameter for the network in (a), by following an incident field at frequency $\omega_p$ as it propagates through the device.  Purple (green) arrows indicate fields propagating in the left (right) arm of the network. Fields are forward transmitted with amplitude and frequency unchanged, but phase shifted by $\pi$.  (c) Backward transmission through the network in (a).  Fields are transmitted with amplitude, frequency, and phase unchanged.}
\label{fig:FullPhasorSchematic}
\end{center}
\end{figure*}

The circulator presented in this paper may be understood in terms of ``synthetic rotation'' created by the active modulation of the circuit, and analyzed with lumped-element circuit theory or an input-output formalism~\cite{kerckhoff:2015}.  Here we provide a complementary explanation of its operation based upon the frequency-domain dynamics of an analogous model system.  

The model is a lumped-element network of multipliers and delays (Fig.~\ref{fig:FullPhasorSchematic}a) which creates the most fundamental non-reciprocal circuit element: a gyrator~\cite{tellegen:1948}. 
In a framework where electromagnetic fields propagate in guided modes into and out of a bounded network at ports, the effect of a network can be described by its scattering matrix element $S_{\mu\nu}$, the ratio of the outgoing field at port $\mu$ to the incident field at port $\nu$~\cite{pozar:2011}. 
Gyrators are linear two-port networks defined by the scattering matrix~\cite{pozar:2011}
\begin{equation}
\bf{S} = \left(\begin{array}{cc} 0 & 1\\ -1 & 0 \end{array}\right).
\label{gyrator}
\end{equation}
In a gyrator, non-reciprocity is manifest in the form of a phase shift.  Fields incident on port $2$ are transmitted with their phase unchanged, but fields incident on port $1$ are transmitted with a $\pi$ phase shift. 

Gyration in the model system arises from the non-commutation of successive translations in frequency and time~\cite{rosenthal:2017}: the multiplying circuits operate as frequency converters, translating an input signal up and down in frequency, and the delays translate fields forward in time.  As frequency and time are Fourier duals, the time-ordering of these translations matters (the two operations do not generally commute).  Transmission through the network thus depends on the propagation direction of the incident signal, breaking Lorentz reciprocity.

To see that non-reciprocity explicitly, frequency-phase diagrams are used to calculate the model's scattering parameters.  The diagrams follow an incident signal at frequency $\omega_p$ as it propagates through the device, tracking its amplitude, frequency, and phase in a frame rotating at $\omega_p$.    

The insets in Fig.~\ref{fig:FullPhasorSchematic}a depict the way that the model system's two constituent elements: multipliers and delays, transform input fields to output fields.  In the multiplying elements, that transformation occurs via multiplication by a bias signal---in this case, $\cos(\Omega t)$.  In the frequency domain, this multiplication creates two sidebands, each detuned from $\omega_p$ by the bias frequency $\Omega$.  Importantly, the phases of these sidebands depend on the phase of the multiplier's bias signal.  We choose a phase reference such that multiplication by $\cos(\Omega t)$ creates two sidebands with the same phase.  

In the delay elements, inputs are transformed to outputs by way of a phase shift. In the rotating frame, a delay of length $\tau = \pi/2 \Omega$ 
advances the phase of spectral components in the upper sideband $\omega_p + \Omega$ by $\pi/2$, and retards the phase of components in the lower sideband $\omega_p - \Omega$ by $\pi/2$. 

With the action of the multiplying and delay elements defined, calculation of the scattering parameters is straightforward.  Forward transmission through the model system is shown in Fig.~\ref{fig:FullPhasorSchematic}b.  A signal incident on port 1 with frequency $\omega_p$ (Fig.~\ref{fig:FullPhasorSchematic}b,~i.) is first divided equally into the network's two arms.  
Fields in both arms encounter a first multiplying element, a delay, a second multiplying element, and are then recombined. 

Critically, the modulation sidebands at $\omega_p \pm 2 \Omega$ created in the network's two arms are $\pi$ out of phase and interfere destructively at the device's output (Fig.~\ref{fig:FullPhasorSchematic}b,~iv).  Conversely, the components at the frequency $\omega_p$ interfere constructively.  Comparison of Fig.~\ref{fig:FullPhasorSchematic}b,~iv. with Fig.~\ref{fig:FullPhasorSchematic}b,~i. shows that the incident signal has been transmitted through the device with its frequency and amplitude unchanged, but its phase shifted by $\pi$.  The scattering parameter $S_{21}$ for the network is therefore $-1$.

The reverse path is traced out in Fig.~\ref{fig:FullPhasorSchematic}c, for a signal incident on the network's second port.  As with forward transmission, destructive interference occurs at $\omega_p \pm 2 \Omega$ (Fig.~\ref{fig:FullPhasorSchematic}c,~iv.).  Likewise, this is accompanied by constructive interference at the frequency $\omega_p$.  Now, however, comparison of Fig.~\ref{fig:FullPhasorSchematic}c,~iv. with Fig.~\ref{fig:FullPhasorSchematic}c,~i. shows that the frequency, amplitude, and phase of the incident signal were unchanged by the network.  Therefore, in contrast to the forward transmission, the backwards transmission is characterized by a scattering parameter $S_{12} = 1$.  The network in Fig.~\ref{fig:FullPhasorSchematic}a is thus described by the scattering matrix of Eq.~(\ref{gyrator}), and forms an ideal gyrator.


The convert-delay-convert process happens simultaneously in both arms of the network.  Consequently, each arm is individually non-reciprocal. Alone, though, a single arm creates unwanted modulation sidebands.
To create an ideal gyrator, two arms, with the bias signals of their multiplying elements separated in phase by $\pi/2$, are connected in parallel. This balanced architecture engineers destructive interference of the spectral components at $\omega_p \pm 2 \Omega$. 

Such a strategy for suppressing the creation of spurious sidebands, which we refer to as ``coherent cancellation,'' may be contrasted with that used in non-reciprocal devices that operate with the parametric coupling of resonant modes in the resolved-sideband limit~\cite{abdo:2014,estep:2014,ranzani:2015,sliwa:2015,lecocq:2017,abdo:2017,bernier:2017,peterson:2017,barzanjeh:2017,metelmann:2017}. 
In that scheme, parametric modulation of a resonant system 
creates sidebands at the parametric drive frequency, and a second resonant mode is used to enhance the density of states at the desired frequency, while simultaneously diminishing it at the undesired frequency. To work in the resolved sideband limit, however, the parametric modulation must be many times the resonant system's linewidth. In microwave frequency implementations, this typically requires GHz modulation tones.  In contrast, the coherent cancellation approach lifts the resolved-sideband constraint, and can therefore be used with lower-frequency control tones.


\section{Superconducting Implementation}
\begin{figure}[!tb]
\begin{center}
\includegraphics[width=1\linewidth]{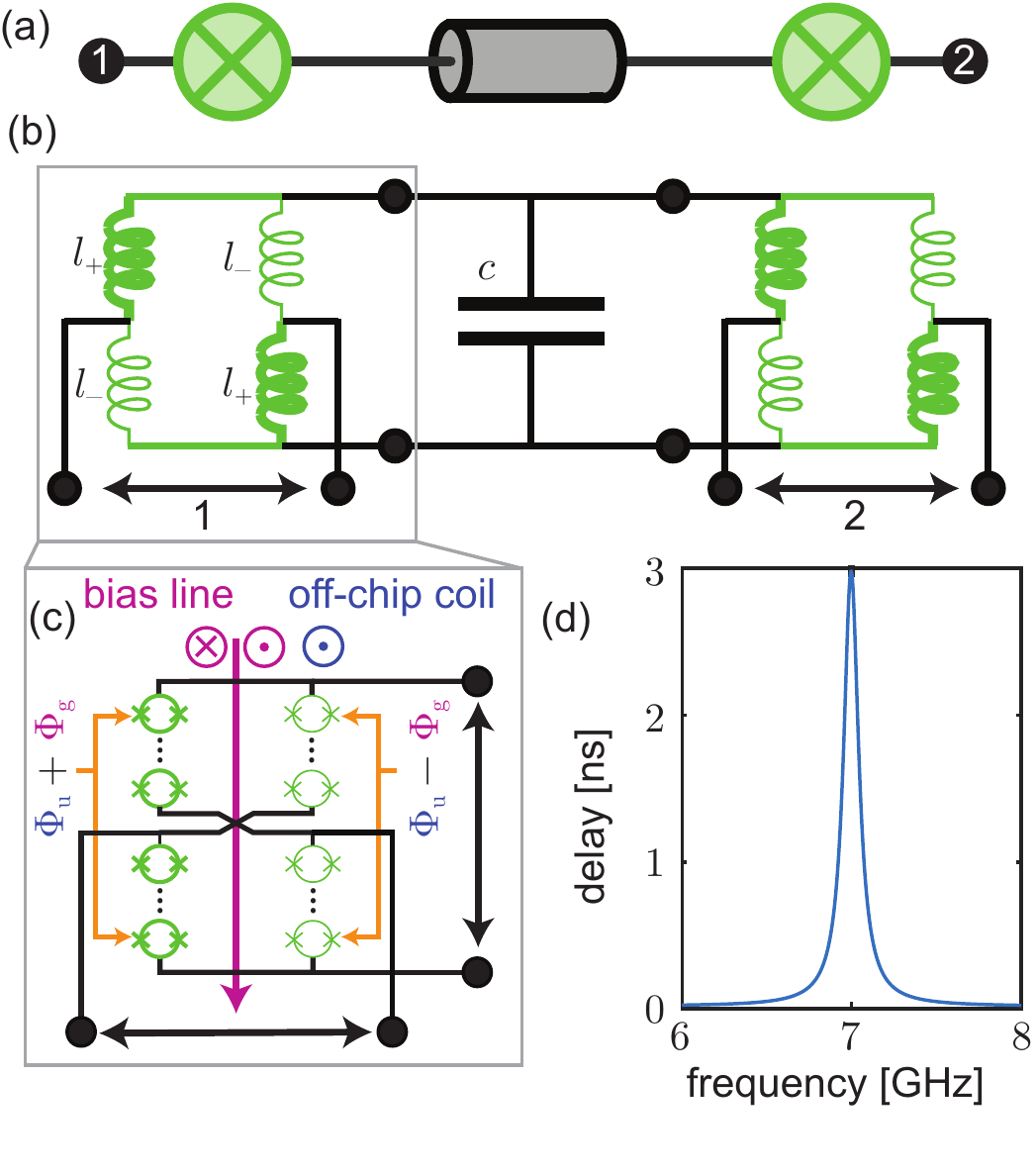}
\caption {
Multiplying elements and delays realized in a superconducting lumped-element circuit. (a) 
The model system (Fig.~\ref{fig:FullPhasorSchematic}a) is constructed from two parallel instances of this network. (b) A lumped-element version of the network in (a), created with capacitors and tunable inductors arranged in a bridge geometry.  
(c) To create an inductive bridge circuit in a superconducting microwave environment, four series-arrays of SQUIDs are arranged in a figure-eight geometry, and tuned with an off-chip magnetic coil producing a uniform flux $\Phi_u$ and an on-chip bias line creating a gradiometric flux $\pm \Phi_g$.  (d) Simulated group delay for the circuit in (b) when its ports are connected to 50 Ohm transmission lines.  The bridge inductors are parametrized according to Eq.~(\ref{l12}), with $c = 1$ pF,  $l_0 = 1$ nH, and $\delta = 0.2$.}
\label{fig:modprim}
\end{center}
\end{figure}

We make use of the unique properties of superconducting circuitry to realize compact on-chip multiplier and delay elements. Specifically, Josephson junctions form widely tunable inductors, while vanishing conductor loss permits on-chip high quality microwave resonators. Fig.~\ref{fig:modprim} shows how a single arm of the model system (Fig.~\ref{fig:modprim}a) is made with a network of capacitors and dynamically tunable inductors (Fig.~\ref{fig:modprim}b).

\subsection{Multiplying elements}

The multiplying elements in the circuit representation are created with reactive bridge circuits, built with two tunable pairs of nominally identical inductors $l_+$ and $l_-$ arranged opposite one-another (gray box in Fig.~\ref{fig:modprim}b). Two differential ports are defined by the left-and-right and top-and-bottom bridge nodes. Importantly, the inductors tune in a coordinated fashion: when one pair of inductors increases, the other pair decreases.  We parametrize this tuning with a base inductance $l_0$ and an imbalance variable $\delta$, by writing
\begin{equation}
l_{\pm} = l_0/(1 \pm \delta).
\label{l12}
\end{equation}  
As the imbalance in the bridge determines its transmission, changing $\delta$ allows the circuit to act as a switch or a multiplying element~\cite{chapman:2016,chapman:2017}.  

The bridge circuit's tunable inductors are realized with series-arrays of superconducting quantum interference devices (SQUIDs), formed by the parallel arrangement of two Josephson junctions.  Arrays are used in place of individual SQUIDs to increase the linearity of the inductors~\cite{kerckhoff:2015}.  The inductance $l$ of an $N$ SQUID array depends on the magnetic flux $\Phi$ that threads through each SQUID~\cite{likharev:1986}:
\begin{equation}
    l = N \frac{\varphi_0}{2 I_0} \left|\sec{\left(\frac{\Phi}{2\varphi_0}\right)}\right| + \mathcal{O}(I/I_0)^2.
    \label{lsquid}
\end{equation}
Here $\varphi_0=\hbar/2e$ is the reduced flux quantum, $I_0$ is the Josephson junction critical current, and the junctions and SQUIDs are assumed to be identical 

To realize the coordinated tuning of inductors described in Eq.~(\ref{l12}), we arrange the SQUID arrays in a figure-eight geometry (Fig.~\ref{fig:modprim}c).  Two flux controls determine the imbalance in the bridge.  First, an off-chip coil threads a uniform magnetic flux $\Phi_u$ through all the SQUIDs.  Second, an on-chip bias line, which bisects the figure-eight, threads a gradiometric flux $\Phi_g$ through the SQUIDs.  SQUIDs on the left side of the bias line therefore experience an overall magnetic flux which is the sum of the uniform and gradiometric contributions, whereas SQUIDs to the right of the line are threaded by the difference of the uniform and gradiometric fluxes.  

When the gradiometric bias line is driven with a sinusoidal current source at frequency $\Omega$, the flux through the SQUIDs varies in time as $\Phi = \Phi_u \pm \Phi_g \cos(\Omega t + \phi)$.  This process creates a bridge of inductors which tune according to Eq.~(\ref{l12}), with a simple sinusoidal variation in the imbalance $\delta = \delta_0 \cos(\Omega t + \phi)$ and a rescaling of the base inductance $l_0$.  App.~\ref{app:KL} describes the mapping between the flux controls $\Phi_u$, $\Phi_g$ and the circuit parameters $l_0$,  $\delta_0$.

\subsection{Delays}


The second primitive needed for the model system is a delay, realized in our circuit with a resonant mode.  Conveniently, the SQUIDs in the bridge circuits are inductive, so the addition of a single capacitor is enough to create a resonance.  This resonance delays fields near its center frequency by a timescale $\tau$ characterized by the inverse of its linewidth.  More quantitatively, when a harmonic field incident on port $\nu$ of a resonant network is scattered to port $\mu$, it acquires a group delay $\tau = d\angle S_{\mu\nu}/d\omega$~\cite{pozar:2011}.  Here $\omega$ is the frequency of the harmonic field, and $\angle S_{\mu\nu}$ is the phase of $S_{\mu\nu}$.  Fig.~\ref{fig:modprim}d shows delay as a function of frequency, for the resonant circuit in Fig.~\ref{fig:modprim}b. 
Fields near the circuit's resonant frequency experience a delay of several nanoseconds.  


Delays realized with resonant modes allow for a deeply sub-wavelength implementation, which is critical for the ``coherent cancellation'' approach.  While these lumped-element delays are necessarily narrower in bandwidth than those created with a length of transmission line, their finite bandwidth is mitigated by the tunable inductance of the bridge circuits, which allows the frequency $\omega_0$ of the resonant delay to be tuned (over several GHz) with the uniform magnetic flux $\Phi_u$.  As the multiplying elements are broadband~\cite{chapman:2017}, this tunability of the delay is inherited by the full circulator.  Likewise, the duration $\tau$ of the delay depends on the imbalance in the bridges, and may be tuned with the gradiometric flux $\Phi_g$, facilitating satisfaction of the requirement that $\tau = \pi/2\Omega$. 


\begin{figure*}[!bht]
\begin{center}
\includegraphics[width=1.0\linewidth]{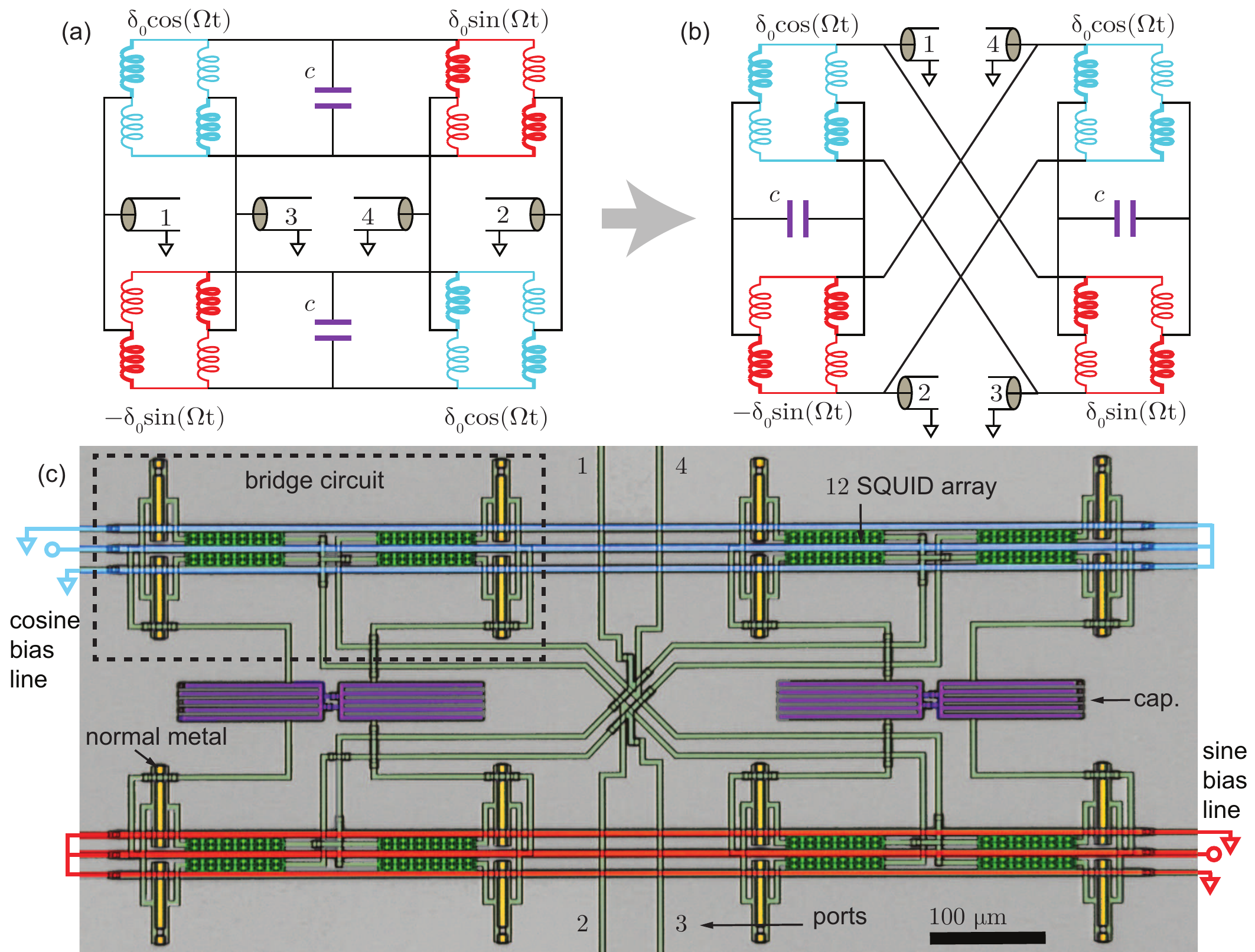}
\caption {
Superconducting realization of a four-port circulator.  (a) Lumped-element schematic of the circulator, constructed with four inductive bridge circuits and two capacitors.  Bias lines (not shown) tune the imbalance $\delta$ of the four bridge circuits dynamically. 
(b) The schematic in (a) redrawn to better match the device layout. 
The rearrangement allows biasing of the bridge circuits with a pair of straight flux-control lines. (c) False-color optical micrograph of the circulator chip. A dashed black box indicates one of the four bridge circuits, created with four 12-SQUID arrays (green) arranged in the figure-eight geometry described in Fig.~\ref{fig:modprim}(c).  Parallel-plate capacitors (purple) provide the capacitance needed for the resonant delay. Bias lines (red and blue) traverse the chip and imbalance the bridge circuits with a gradiometric flux.}
\label{fig:ChipPhoto}
\end{center}
\end{figure*}

Tuning of the resonant delay takes a simple form when expressed in terms of the circuit parameters $l_0$ and $\delta_0$.  When two of the arms in Fig.~\ref{fig:modprim}b are combined in parallel to create the fully assembled circuit shown in Fig.~\ref{fig:ChipPhoto}a, the resonant delay occurs at the frequency~\cite{kerckhoff:2015}
\begin{equation}
    \omega_0 = \sqrt{\frac{4 - \delta_0^2}{2 l_0 c}},
    \label{w0}
\end{equation}
and its duration $\tau$ is approximately the inverse of the resonant mode's linewidth,
\begin{equation}
    \tau \approx \frac{8 Z_0 c}{\delta_0^2}.
    \label{tau}
\end{equation}
Here $Z_0$ is the characteristic impedance of the surrounding transmission lines.


\subsection{Circulator}
Assembly of a superconducting version of the full circuit requires the parallel combination of two arms like the one shown in Fig.~\ref{fig:modprim}b. Fig.~\ref{fig:ChipPhoto}a shows a lumped element schematic of the complete network. When ports 1 $\&$ 3 and 2 \& 4 are driven differentially, the circuit in Fig.~\ref{fig:ChipPhoto}a creates a gyrator that functions in the same way as the model system (Fig.~\ref{fig:FullPhasorSchematic}a).  If instead, however, ports are defined by comparison of voltages to a common ground (as shown in Fig.~\ref{fig:ChipPhoto}a), the circuit forms a reconfigurable four-port circulator, with (ideal) clockwise scattering matrix:
\begin{eqnarray}
\bf{S} &=& \left(\begin{array}{cccc} 0 & 1 & 0 & 0 \\ 0 & 0 & 1 & 0 \\ 0 & 0 & 0 & 1 \\ 1 & 0 & 0 & 0 \end{array}\right),
\label{circulator}
\end{eqnarray}
or counterclockwise scattering matrix $\bf{S}^T$.  

The transformation between gyrator and circulator can be understood in the following way: driving any one of the device's four ports involves simultaneously exciting the common and differential modes of the circuit.  The gyrating differential mode is non-reciprocal, whereas the prompt scattering of the non-resonant common mode is reciprocal.  The interference of these two scattering processes results in circulation~\cite{kerckhoff:2015}.

For fabrication, the circuit is laid-out as illustrated in Fig.~\ref{fig:ChipPhoto}b.  This rearrangement leaves the connectivity of the circuitry unchanged, but allows two pairs of parallel flux-control bias lines to bisect the four inductive bridges. Fig.~\ref{fig:ChipPhoto}c shows a false-color optical micrograph of the device, laid out in this way.  



The SQUID arrays that comprise the circuit's multiplying elements are colored dark green in the micrograph and oriented horizontally.  Capacitors, to make the resonant delays, are realized in a parallel-plate, metal-insulator-metal geometry, and colored purple.  Finally, normal metal (Au) is used in the layout to break supercurrent loops which can trap flux.  These sections are colored gold, and have a resistance of approximately $10$ milliohms, which increases insertion loss by 0.1 dB.  Further details on the layout design considerations are provided in App.~\ref{app:engineering}.

\section{Experimental Results}

\begin{figure}[htb]
\begin{center}
\includegraphics[width=1\linewidth]{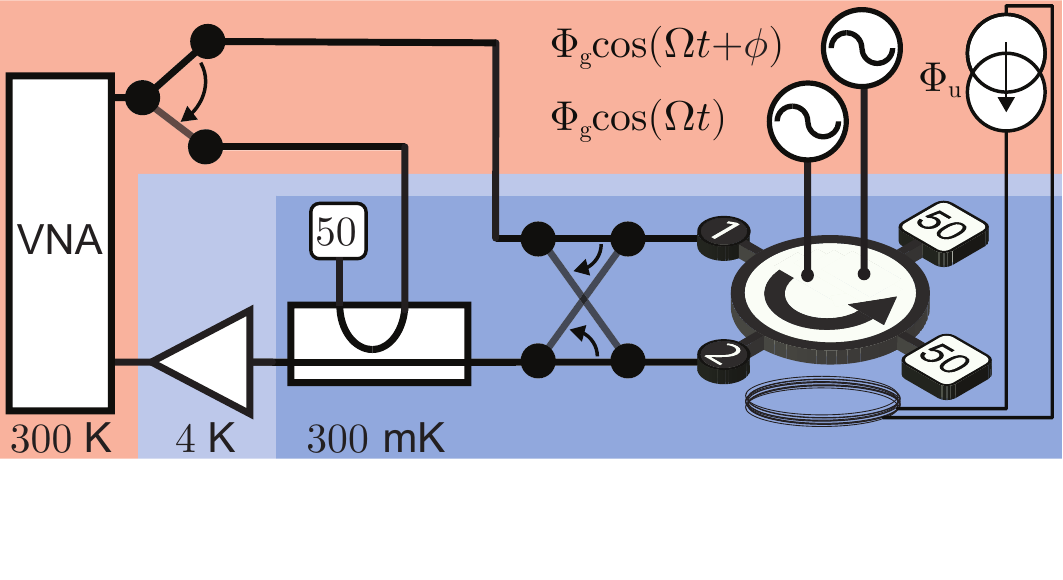}
\caption {Simplified experimental schematic for circulator measurements in a $^3$He cryostat (attenuation, filtering, and isolation omitted) made with a vector network analyzer (VNA).  Two switches and a directional coupler allow measurement of four of the circulator's scattering parameters.
}
\label{fig:nakedexptschem}
\end{center}
\end{figure}

To test the device, two of its four ports are terminated in $50$ Ohm loads and the circuit is mounted at the base of a $^3$He cryostat.  A schematic of the experimental setup is shown in Fig.~\ref{fig:nakedexptschem}.  Two switches and a directional coupler allow for measurement of the four accessible scattering parameters.  

\begin{figure}[!thb]
\begin{center}
\includegraphics[width=1\linewidth]{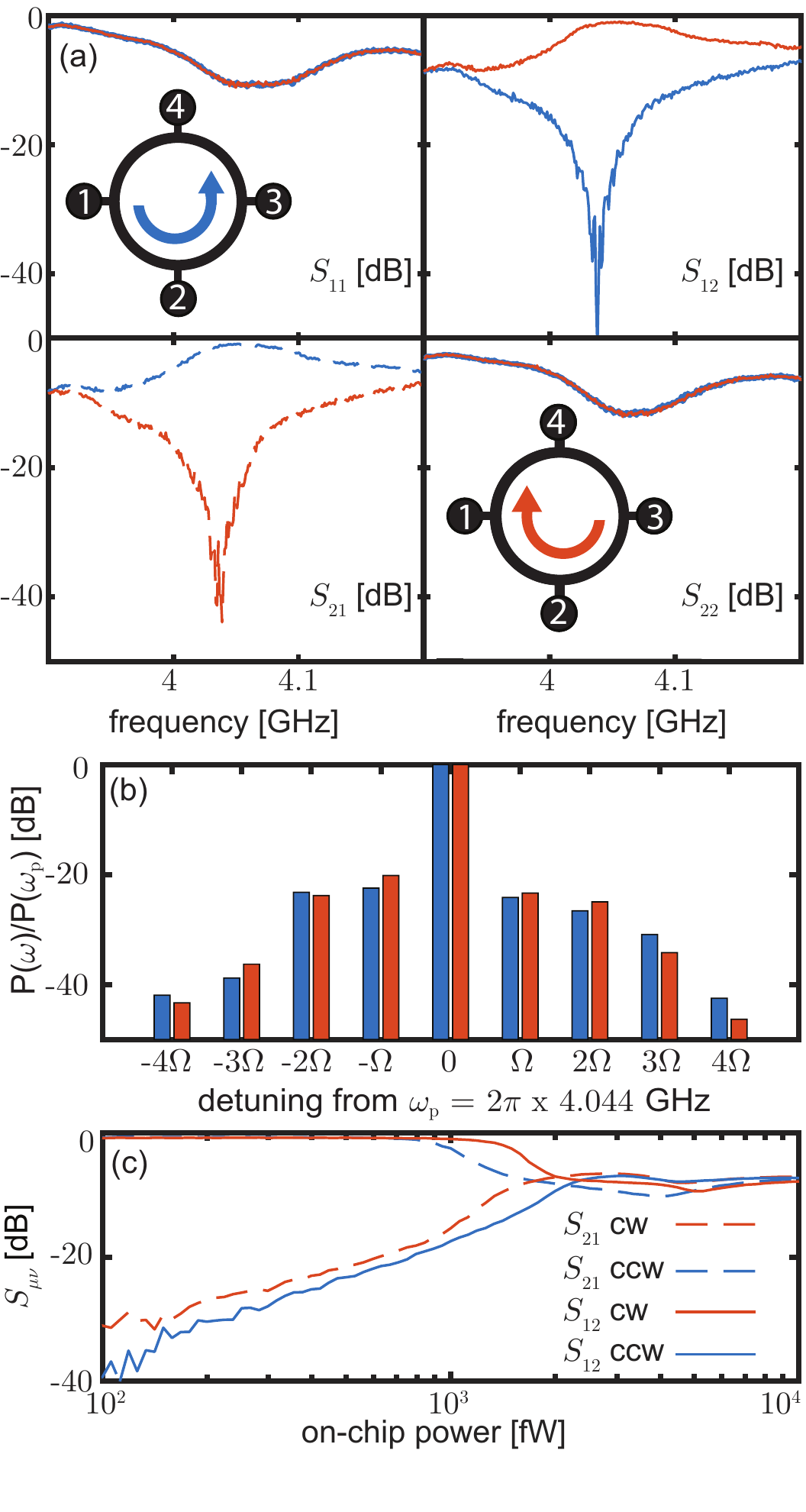}
\caption {
Performance of an on-chip superconducting circulator tuned to operate near 4 GHz.  (a) Frequency dependence of four of the circulator's 16 scattering parameters, when configured as a counterclockwise circulator (blue traces) and a clockwise circulator (orange traces). (b) Transmission spectrum of the circulator at $\omega_p = 2 \pi \times 4.044$ GHz, measured at frequencies $\omega_p \pm m \Omega$ with $m$ a positive integer.  Spectral components are normalized by the power transmitted at $\omega_p$.  Spurious sidebands are suppressed by approximately $20$ dB. (c) Transmission as a function of probe power, with probe frequency fixed at $\omega_p = 2 \pi \times 4.044$ GHz.  1 dB compression occurs around 1 pW.}
\label{fig:Sparameters}
\end{center}
\end{figure}

App.~\ref{sec:tuneup} describes the tune-up procedure, which involves choice of the modulation frequency $\Omega$ and the delay $\tau$, as well as selection of the phases $\phi_{\textrm{cw}}$ and $\phi_{\textrm{ccw}}$  that best accomplish clockwise and counterclockwise circulation. 
Fig.~\ref{fig:Sparameters} shows the results of this process, when the device is tuned to operate near 4 GHz.   Four of the device's sixteen scattering parameters are plotted in Fig.~\ref{fig:Sparameters}a, for both the clockwise and counterclockwise operation modes.  Different ports were probed in a separate cooldown, with similar results. (Calibration of network parameter measurements is discussed in App.~\ref{app:Cal}.)

In the transmission measurements (top right and bottom left plots), high transmission ($> -1$ dB) and robust isolation ($> 20$ dB) are observed in a $50$ MHz window around $4.044$ GHz.  These features are approximately coincident with $-11$ dB dips in the reflection measurements (top left and bottom right plots). 
Together, power collected in the transmission and reflection measurements account for 90\% of the injected signal power.

To determine if the remaining power is dissipated or scattered to other frequencies, a spectrum analyzer is used to measure the transmission of the circulator at sidebands of the modulation frequency $\omega_p \pm m \Omega$.  
Fig.~\ref{fig:Sparameters}b shows the power of these spectral components, relative to the transmitted spectral component at $\omega_p$.  The device suppresses spurious sidebands by more than 20 dB.  The spectral purity of the output---in particular, the suppression of spectral components at $\omega_p \pm 2 \Omega$---is a testament to the high-degree of symmetry in the circuit. 
From this measurement, we conclude that the remaining 10\% of input power is dissipated into heat or other radiation modes.

Finally, Fig.~\ref{fig:Sparameters}c displays the dependence of clockwise and counterclockwise transmission on the power of the probe signal.  Fixing the probe frequency at $\omega_p = 2 \pi \times 4.044$ GHz, the measurement is repeated for both clockwise and counterclockwise operation. In both cases, 1 dB compression of the transmitted signal occurs at input powers around 1 pW.  As the input power approaches this value, we also observe a degradation in the circulator's isolation, which drops below 20 dB at a power again roughly equal to 1 pW.  In analogy with the 1 dB compression point, we refer to this power as the 20 dB expansion point of the circulator.  Expressed in terms of photon number, this linearity allows the circulator to process over $10^3$ photons per inverse of its bandwidth.

For reference, the typical power in a microwave tone used for dispersive readout~\cite{blais:2004} of a superconducting qubit is between $100$ and $1000$ aW (few photon level)~\cite{riste:2013}.  The three orders of magnitude that separate this power scale from the 1 dB compression and 20 dB expansion points of the device are critical for one attractive application of a monolithic superconducting circulator: on-chip integration with a quantum-limited reflection amplifier, such as a Josephson parametric amplifier.  The high power handling of the circulator allows it to route qubit readout tones even after they reflect off a Josephson parametric amplifier and are amplified by 20 dB.

\begin{figure}[tb]
\begin{center}
\includegraphics[width=1\linewidth]{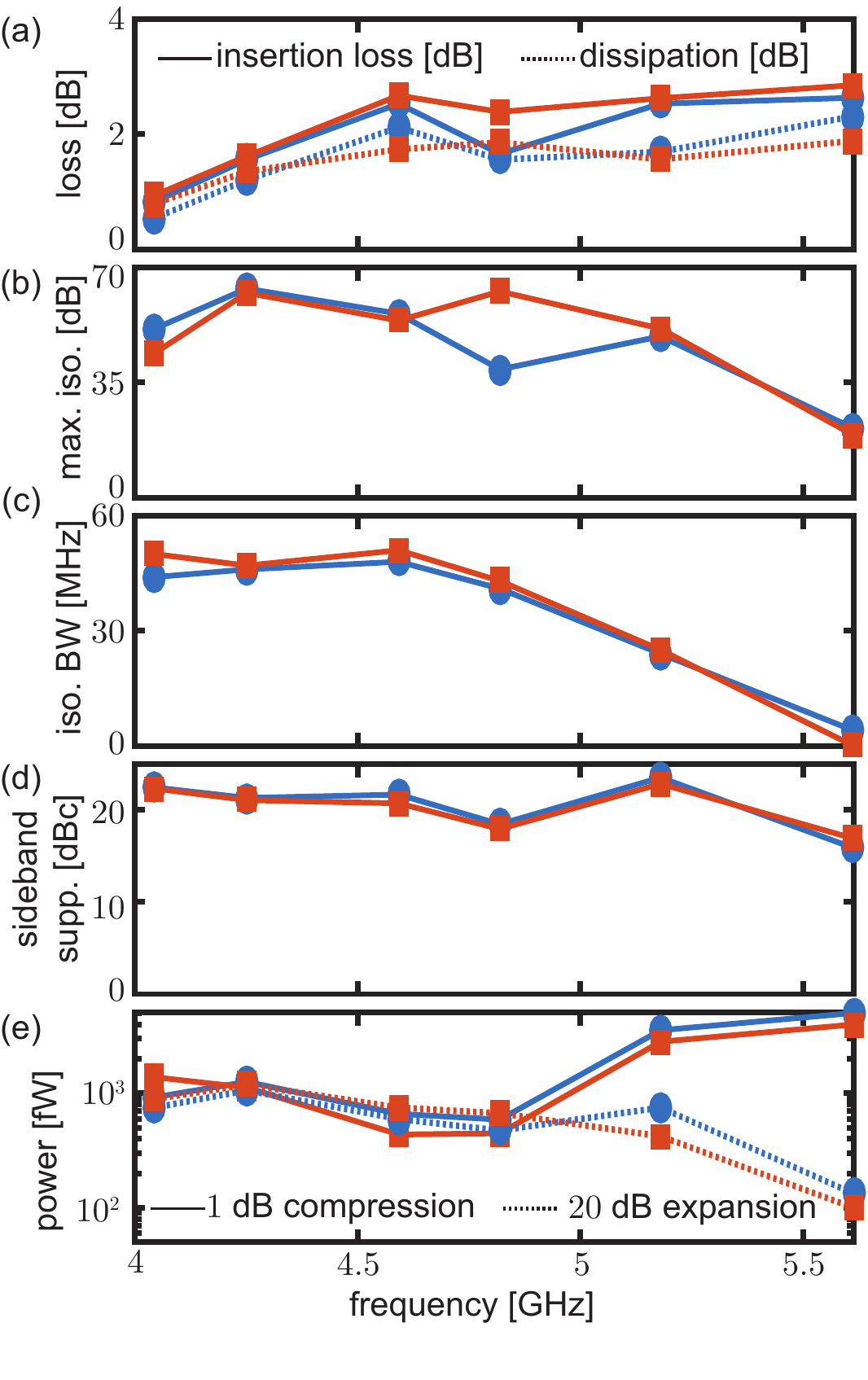}
\caption {
Performance of a widely tunable on-chip circulator.  The insertion loss (a, solid lines), dissipation (a, dotted lines), maximum isolation (b), isolation bandwidth (c), sideband suppression (d), and power handling (e) are plotted at operation frequencies between 4 and 6 GHz, when the device is configured as a clockwise (orange rectangles) or counterclockwise (blue circles) circulator.  Isolation bandwidth refers to the bandwidth over which isolation exceeds 20 dB.  Sideband suppression refers to the contrast between transmitted spectral components at the input frequency $\omega_p$ and the largest spurious sideband of the form $\omega_p \pm n \Omega$.  In (e), dotted lines indicate the device's 1 dB compression point.  Solid lines indicate the 20 dB expansion point, where isolation drops below 20 dB.}
\label{fig:metadata}
\end{center}
\end{figure}

To demonstrate the circulator's tunability, we operate the device at a variety of frequencies between 4 and 6 GHz and repeat the measurements shown in Fig.~\ref{fig:Sparameters}. 
Fig.~\ref{fig:metadata} summarizes the performance of the device across this tunable range.  
Insertion loss is shown in Fig.~\ref{fig:metadata}a.  Transmission is greatest at the lowest frequency, and decreases with frequency until it approaches -3 dB.  
We attribute this trend to the geometric inductance present in the circuit, which limits the degree to which the bridges can be imbalanced.  This inhibits impedance matching and reduces the degree to which the resonant differential modes are over-coupled.  At lower operation frequencies, the Josephson inductance comprises a greater fraction of the bridge's total inductance, mitigating this effect.

This interpretation is supported by the power dissipation that we estimate at each operation frequency (Fig.~\ref{fig:metadata}a, dashed lines), computed as the sum of the reflection and transmission coefficients $-10\log{\left(R^2 + T^2\right)}$. (Power transmitted to sidebands of the modulation frequency is suppressed by over 20 dB, and is therefore neglected in this accounting).  Reflections, visible in the discrepancy between insertion loss and transmission, are larger at higher frequencies, where the role of geometric inductance is more pronounced.  Dissipation is also greater at higher operation frequencies, where the external coupling of the resonant mode is lower.  One of the principal dissipation sources in the circuit is the dielectric loss of SiO$_2$, which results in a frequency-dependent loss that we estimate ranges from 0.3 to 0.8 dB.  

The circulator's maximum isolation is plotted in Fig.~\ref{fig:metadata}b.  Below 5.5 GHz, isolation exceeds 35 dB for both device configurations. Critically, isolation is achieved over a bandwidth of several tens of MHz, much greater than the bandwidths typical for strongly-coupled cavity ports in dispersive qubit readout, which range up to several MHz~\cite{kelly:2015,ofek:2016,hacohen:2016}.  Fig.~\ref{fig:metadata}c shows the frequency interval over which the isolation exceeds 20 dB.  

It should be noted this isolation is achieved concurrent with the performance shown in the rest of Fig.~\ref{fig:metadata}: all specifications are measured at two fixed operation phases, which realize clockwise and counterclockwise circulation. To select these operation phases in a quantitative manner, we write a cost function to simultaneously balance the benefits of low insertion loss, high isolation, and broad bandwidth, for both clockwise and counterclockwise operation. 
Ultimately, different applications will prioritize the relative importance of these specifications in different ways, allowing trade-offs in performance specifications, for example, between insertion loss and isolation. Similarly, if the device's reconfigurability is not needed, performance will generally exceed that shown in Fig.~\ref{fig:metadata}.  

Fig.~\ref{fig:metadata}d characterizes the spectral purity of transmitted fields at each operation frequency.  It shows the size of the largest spurious sideband, (relative to the power transmitted at the probe frequency), which we call the sideband suppression.  
Spurious sidebands are strongly suppressed across the operation range, typically by about 20 dB.

Lastly, Fig.~\ref{fig:metadata}e shows how the power-handling of the circulator depends on the operation frequency.  Frequencies between 4 and 5 GHz have 1 dB compression points and 20 dB expansion points around 1 pW. 

\section{Conclusion and Outlook}
In this work we realize the on-chip superconducting circulator proposed in Ref.~\cite{kerckhoff:2015}.  Lorentz reciprocity is broken in the circuit with sequential translations in frequency and time, which we show with a simple model system composed of just two components: multiplying elements and delays.  We describe how both elements can be created in a cryogenic microwave environment, and then characterize the performance of a circulator built from these components.  We observe low insertion loss and over $20$ dB of isolation over a bandwidth of approximately 50 MHz.  The device is linear with respect to input power for fields up to $1$ pW in power, and its transmission spectrum is spectrally pure, in the sense that spurious harmonics created by the device's RF control tones are suppressed by more than $20$ dB. Finally, we demonstrate that all of these performance specifications can be achieved over a tunable operating range approaching $2$ GHz, and in clockwise or counterclockwise configurations. 

As the device is controlled with radio frequency tones (which are a) easily phase-locked and b) require none of the limited high-bandwidth transmission lines in a dilution refrigerator), and as it is orders of magnitude more compact than commercial ferrite circulators, this superconducting circulator is a scalable alternative to signal routing with ferrite junction circulators. We estimate that with superconducting twisted pairs carrying the low-frequency control tones, $10^3$ of these circulators could be operated in a single dilution refrigerator (see App.~\ref{app:filtering}).


Looking forward, the work suggests several immediate extensions.  In a future design, layout changes could improve device performance: dielectric loss can be reduced with the use of low-loss dielectrics like amorphous silicon~\cite{lecocq:2017} or interdigitated capacitors.  Similarly, dividing the power in the gradiometric flux lines off-chip and delivering them on-chip in four dedicated bias lines removes layout constraints, and enables the design of a circuit with approximately half the geometric inductance. Even with the device's existing performance, another obvious extension is on-chip integration of the circulator with a quantum-limited amplifier, for measurements of added noise.  Finally, the essential concept of frequency conversion and delay can be adapted to a lossless and broadband design, using non-resonant delays~\cite{doerr:2011,yang:2014,rosenthal:2017}.  Prospects for such a device are extremely attractive given the high power-handling of these SQUID-array based devices, as their integration with a broadband low-noise amplifier~\cite{macklin:2015} could enable scalable frequency-domain multiplexing of many-qubit systems with near-unit measurement efficiency.

\section*{Acknowledgments}
This work is supported by the ARO under contract W911NF-14-1-0079 and the National Science Foundation under Grant Number
1125844.  

\appendix
\section{Flux control}
\label{app:KL}
The proposal in Ref.~\cite{kerckhoff:2015} analyzes a lumped-element model of the circulator, formed with dynamically tunable inductors parametrized according to Eq.~(\ref{l12}).  The following relations connect that parametrization with the experimental flux-control parameters $\Phi_u$ and $\Phi_g$~\cite{kevinthesis}:
\begin{eqnarray}
\delta_0 &=& -2\tan(\alpha) \frac{J_1(\beta)}{J_0(\beta)} + \mathcal{O}(\beta^2), \nonumber \\
l_0 &=& N \frac{\varphi_0}{2 I_0} \frac{1}{\cos(\alpha) J_0(\beta)} + \mathcal{O}(\beta^2),
\label{fluxmapping}
\end{eqnarray}
with
\begin{eqnarray}
\alpha &\equiv& \pi\frac{\Phi_u}{\Phi_0}, \nonumber \\
\beta &\equiv& \pi\frac{\Phi_g}{\Phi_0},
\label{alpha}
\end{eqnarray}
and $J_n$ the $n^\textrm{th}$ Bessel function of the first kind.

\section{Tune-up procedure}
\label{sec:tuneup}

\begin{figure}[htb]
\begin{center}
\includegraphics[width=1\linewidth]{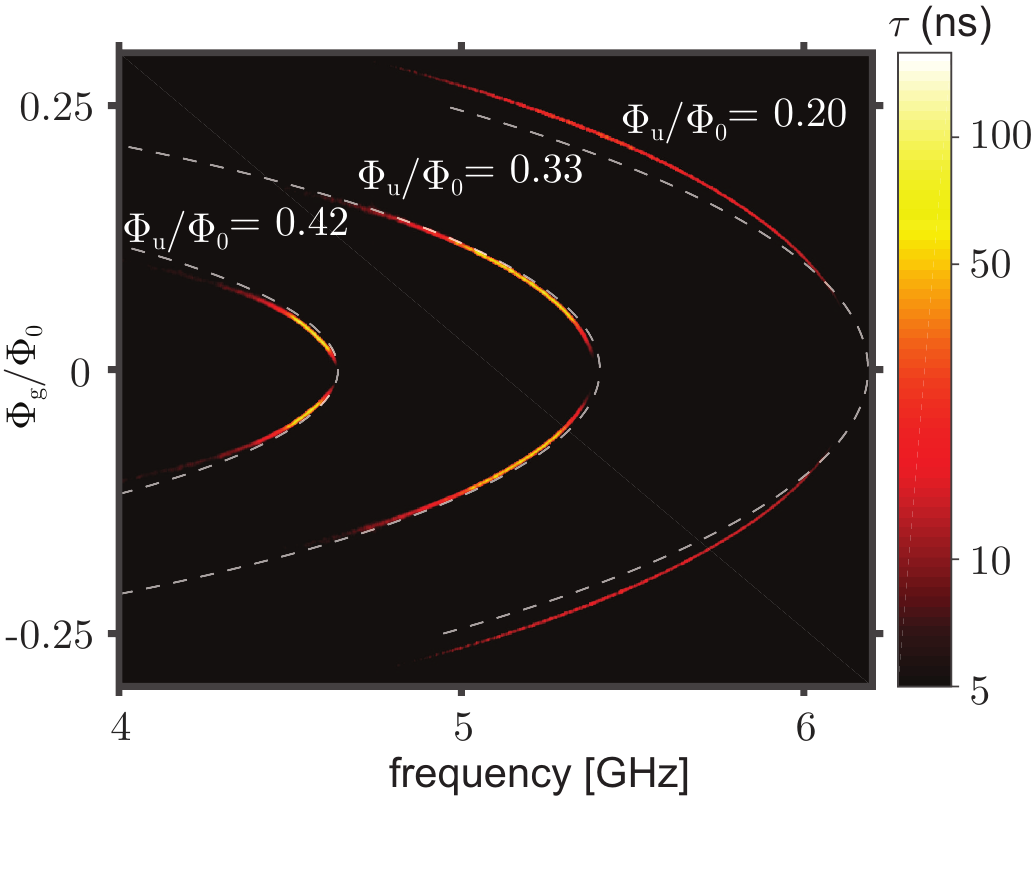}
\caption {Measurements of the circulator's group delay $\tau$ (color, log scale) as a function of the probe frequency and a static gradiometric flux $\Phi_u$ applied to all four of the bridge circuits. The duration and center frequency of the resonant delay depend on the uniform and gradiometric flux, allowing the circulator's operation frequency to be tuned between 4 and 6 GHz. Dashed gray lines are predictions of Eq.~(\ref{w0}) which use the mapping in App.~\ref{app:KL}.  To account for geometric inductance in the circuit (which is not present in the model), the dashed lines are calculated with effective uniform fluxes $\tilde{\Phi}_u/\Phi_0 = 0.38$, $0.33$, and $0.28$ chosen to match the frequency of the measured and predicted delays when $\Phi_g = 0$.
}
\label{fig:delaymeas}
\end{center}
\end{figure}

Three straightforward steps are required to prepare the circulator for operation.  First, the frequency of the resonant delay $\omega_0$ is tuned to the desired operation frequency.  Second, the duration of the resonant delay $\tau$ is set to $\pi/2 \Omega$.  Finally, the phase difference $\phi$ between the gradiometric flux control drives is set to $\pm\pi/2$.  


We illustrate the first two of these steps in Fig.~\ref{fig:delaymeas}, which shows in color the group delay $\tau$ acquired during transmission through the device at different probe frequencies $\omega_p$, and for different values of a static gradiometric flux $\Phi_g$ applied to all four of the inductive bridges. The measurement is shown for three different values of the uniform flux $\Phi_u$.

Two features are immediately evident in the data.  First, the resonant nature of the delay is clear: for fixed values of $\Phi_u$ and $\Phi_g$, fields at most probe frequencies are off-resonance and their group delays are less than 5 ns, as visible in the black background of the color plot.  Against this background, three arches are visible, which show the resonant delay tuning with $\Phi_g$ for the three measurements at distinct $\Phi_u$.  The shapes of these arches are qualitatively captured by the theoretical predictions in dashed gray lines, which are made with Eq.~(\ref{w0}) and the relations in App.~\ref{app:KL} that map $\Phi_g$ \& $\Phi_u$ to $\delta_0$ \& $l_0$.  

Second, when $|\Phi_g|$ approaches $0$, the group delay vanishes.  A gradiometric bias with magnitude much less than $\Phi_0 =2 \pi \varphi_0$ results in approximately balanced bridges ($\delta \ll 1$).  As the external-coupling of the resonant mode depends on $\delta^2$ (Eq.~(\ref{tau})), balanced bridges result in under-coupled resonant modes, which strongly attenuates transmission through the resonant differential modes.  (The internal quality factor of the circuit is estimated to be 400 when the resonant delay is tuned to 5 GHz.) Power is still transmitted through the non-resonant common mode, but without acquiring resonant delay.  


As circulation bandwidth scales with the linewidth of the resonant delay, for the measurements in this paper we operate the device with a relatively brief delay on the order of several nanoseconds, with $\Omega = 2 \pi \times 120$ MHz. This choice has the additional benefit of reducing the influence of internal losses by keeping the circuit strongly over-coupled. 



\begin{figure}[!thb]
\begin{center}
\includegraphics[width=1\linewidth]{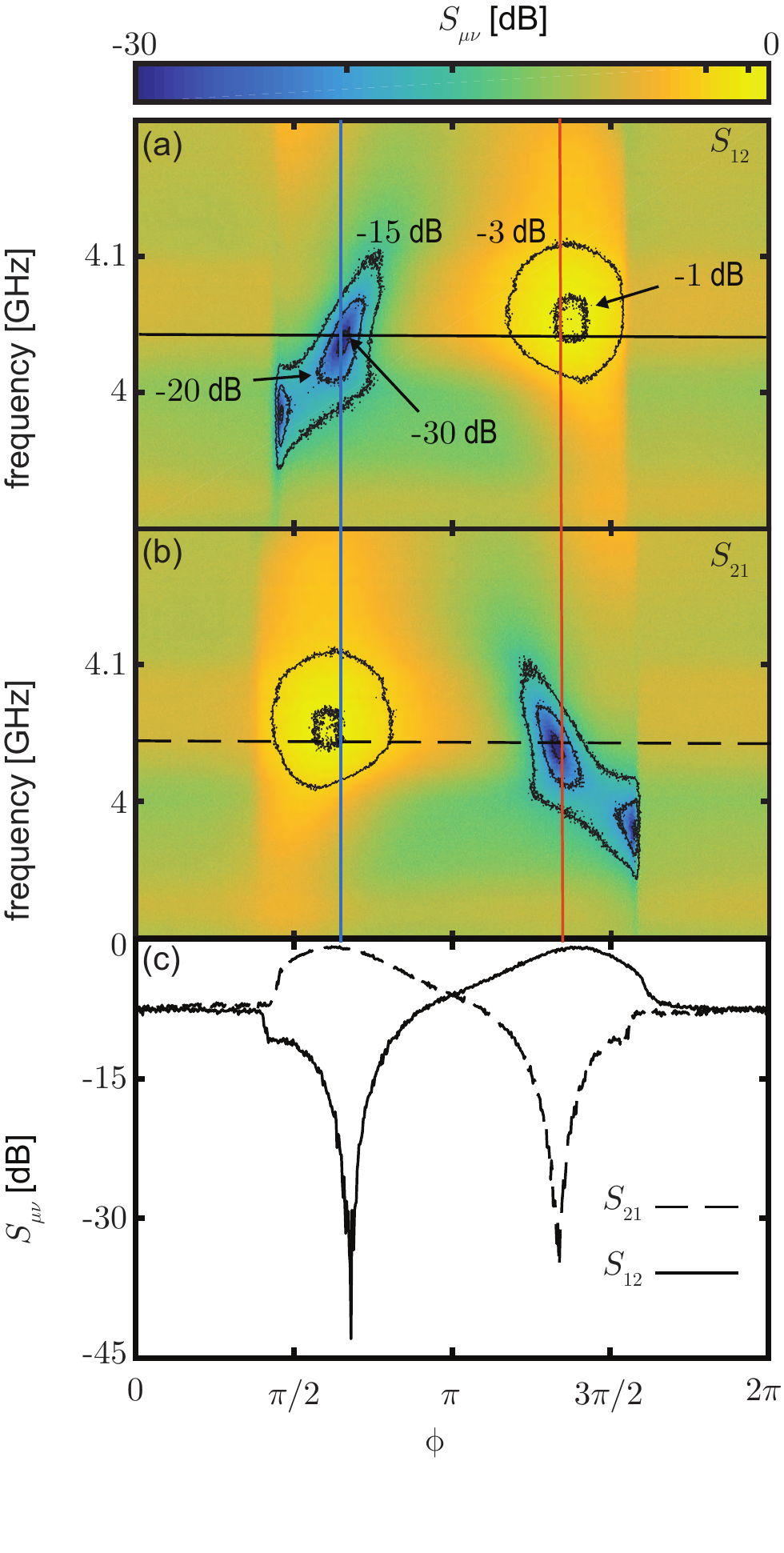}
\caption {
Measurements of a dynamically reconfigurable circulator.  The phase $\phi$ between the gradiometric flux lines determines the direction of circulation.  Counterclockwise (a) and clockwise (b) transmission as a function of the probe frequency and the phase $\phi$.  
Vertical and horizontal lines indicate the location of linecuts plotted in (c) and Fig.~\ref{fig:Sparameters}(a).  (c) Counterclockwise (dashed) and clockwise (solid) transmission at probe frequency $\omega_p = 2 \pi \times 4.044$ GHz.}
\label{fig:phasesweep}
\end{center}
\end{figure}

The final step of the tune-up procedure involves selection of the relative phase $\phi$ between the gradiometric flux controls.  Fig.~\ref{fig:phasesweep} shows a sweep of $\phi$ when the resonant delay $\omega_0$ is set near $2 \pi \times 4$ GHz and the duration of the delay $\tau$ is fixed at several ns with $\Omega = 2 \pi \times 120$ MHz.  The color scale in Fig.~\ref{fig:phasesweep}a shows the magnitude of $S_{21}$ as a function of this phase and the probe frequency: $S_{12}$ is shown in Fig.~\ref{fig:phasesweep}b.

The color plots in Fig.~\ref{fig:phasesweep} reveal two regions of parameter space in which operating points can be chosen. At these phases, the insertion loss is less than 1 dB and the isolation exceeds 30 dB.  They can therefore be interpreted as the phases which realize a clockwise or counterclockwise circulator. 
To illustrate this, Fig.~\ref{fig:phasesweep}c shows frequency linecuts at $4.044$ GHz 
from both transmission measurements.  Importantly, the linecuts show that high transmission in the counterclockwise (clockwise) direction is accompanied by strong isolation in the clockwise (counterclockwise) direction. 
They also illustrate how toggling the phase $\phi$ allows dynamical reconfiguration of the device's sense of circulation. Interestingly, one can see that the strongest non-reciprocity is observed at phases near but distinct from the expected operating points at $\pi/2$ and $3\pi/2$.  App.~\ref{app:imperfections} describes how geometric inductance in the circuit causes this discrepancy.  

\section{Calibration of network parameter measurements}
\label{app:Cal}
\subsection{Transmission calibration}
To remove the gain of the measurement chain in transmission measurements, a bypass switch is mounted at the base of the cryostat, which routes fields through a 5 cm SMA cable instead of the circulator.  We also use dedicated through measurements, (made in a separate cooldown) in which the circulator chip is exchanged for a like-sized circuit board traversed by a single 50 Ohm transmission line.  Using these techniques, the reference plane for transmission measurements is moved (approximately) to the edge of the chip.
\subsection{Reflection calibration}
To remove the gain $G$ from reflection measurements, we measure the reflection $R_{\textrm{bal}}$ off the circulator when no bias current is applied to the on-chip bias lines.  In this unbiased state, all four inductor bridges are balanced, and the reflection coefficient $\Gamma_\textrm{bal}$ is the diagonal entry in each row of the balanced scattering matrix $\textbf{S}_\textrm{bal}$.  The matrix $\textbf{S}_\textrm{bal}$ may be calculated by substituting the lower-right block matrix of Eq.~(8) in Ref. 19 into Eq.~(19) of that reference.  This procedure yields
\begin{equation}
    \Gamma_\textrm{bal} = \frac{i \omega l + 2 Z_0}{i \omega l - 4 Z_0}.
    \label{gamma}
\end{equation}
As
\begin{equation}
R_{\textrm{bal}} = G \Gamma_\textrm{bal},
\end{equation}
and the gain $G$ of the reflection measurement chain is assumed to be independent of the circulator's state, the reflection coefficient $\Gamma_\textrm{op}$ at arbitrary operation points is related to the measured reflection $R_\textrm{op}$ by

\begin{equation}
    \Gamma_\textrm{op} = \frac{R_\textrm{op}}{G} = \Gamma_\textrm{bal} \frac{R_\textrm{op}}{R_\textrm{bal}}.
\end{equation}

To account for geometric inductance in the bridges, the inductance $l$ in Eq.~(\ref{gamma}) is estimated using measurements of the unbalanced circuit's resonant frequency, the capacitance design value of 1~pF, and Eq.~(\ref{w0}).

\subsection{Calibration of group delay}
Preparing the circulator for operation requires correctly setting the duration $\tau$ of the resonant delay.  Measurements of the circulator's group delay are used for this purpose.  To separate the non-resonant delays of the finite-length measurement chain from the resonant delay $\tau$, we multiply the measured transmission data by $e^{i \omega \tau_d}$, where $\tau_d = 62$ ns is the time required for an off-resonant microwave field to propagate through the measurement chain.  In the absence of circuit resonances, this multiplication makes the phase of the transmission flat as a function of frequency, zeroing the group delay.  



\section{Circuit layout}
\label{app:engineering}
The circuit discussed in this work presents several design challenges, some of which are specific to superconducting circuits.

\subsection{Capacitor design}

\begin{figure}[!thb]
\begin{center}
\includegraphics[width=1\linewidth]{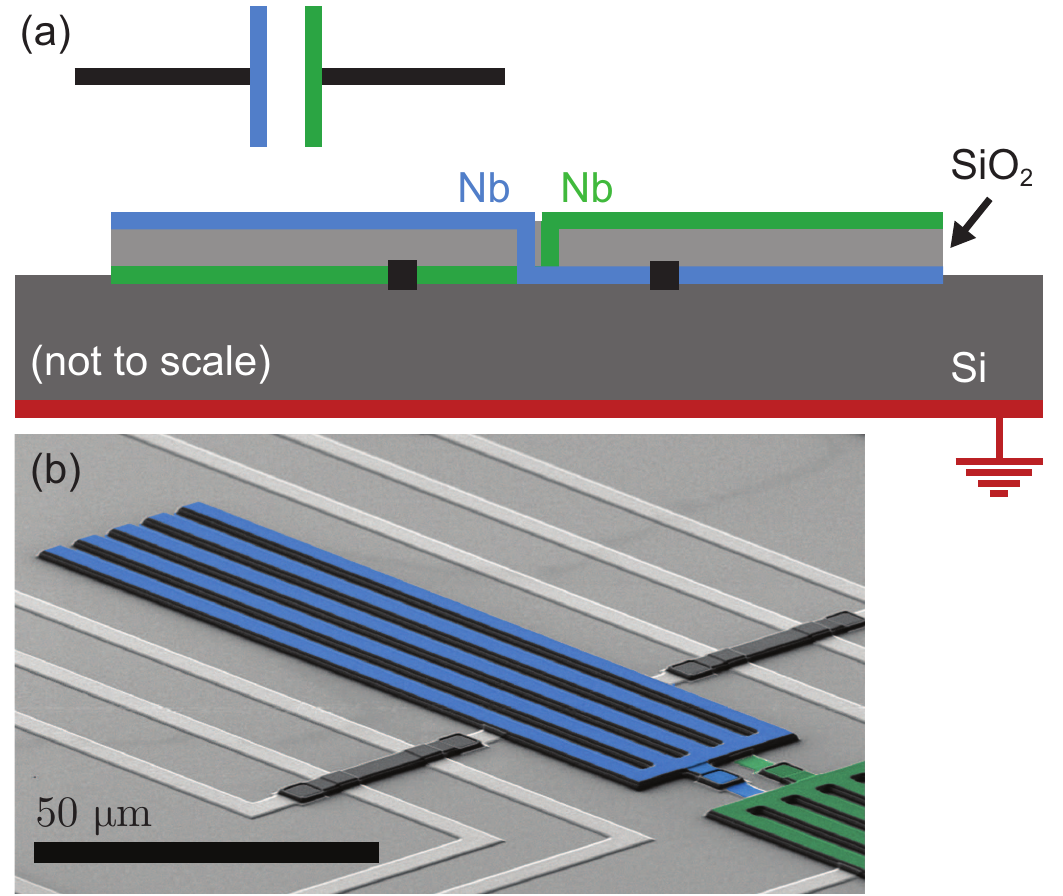}
\caption {
(a) Schematic illustrating the symmetrization of the parallel plate capacitors used in the device.  To create a capacitor $c$, two capacitors with half the desired capacitance are connected in parallel, such that the upper (lower) plate of the first capacitor is galvanically connected to the lower (upper) plate of the second.  This procedure gives each side of the capacitor the same parasitic capacitance to the ground plane. (b) False-color scanning electron microscope image showing one of the $c/2$ capacitors used in the device (blue), and the joint connecting it to another $c/2$ capacitor (green), as described in (a).  The plates in both capacitors are formed from narrow niobium strips of width $w = 5$ \micron to prevent trapping flux vortices~\cite{martinis:2004}.}
\label{fig:CapacitorDesign}
\end{center}
\end{figure}

To realize the capacitors $c$ in the circulator's lumped element representation (Fig.~\ref{fig:ChipPhoto}a), we layout parallel-plate capacitors in a metal-insulator-metal geometry with Nb plates sandwiched around the dielectric SiO$_2$ (Fig.~\ref{fig:CapacitorDesign}a).  In the frequency range of 4 to 8 GHz, roughly pF capacitances are required to create capacitor-impedances near 50 ohms.  Making a pF capacitor with SiO$_2$ in the Nb trilayer process requires capacitor plates that are roughly 100 \micron on a side---large enough to trap magnetic flux vortices when cooled through Niobium's superconducting transition temperature $T_c$ in earth's magnetic field~\cite{martinis:2004}. 

To avoid trapping flux vortices, we pattern slots in the capacitor electrodes, such that the Nb strips that form the electrodes never exceed a width $w \ll$ 100 \micron. This ability to suppress vortices in non-zero magnetic fields is important for our layout, as the circulator is actuated with flux controls which can be spoiled by a static and unremovable flux gradient. Choosing $w = 5$ \micron ensures that the capacitor electrodes trap no magnetic flux vortices when the capacitor is cooled through $T_c$ in our experiment's modestly shielded magnetic environment. 



To further symmetrize the circuit, each parallel-plate capacitor is then divided into two capacitors of capacitance $c/2$, and connected in parallel, such that the upper plate of the first (second) capacitor is galvanically linked to the lower plate of the second (first) capacitor (Fig.~\ref{fig:CapacitorDesign}b).  This procedure gives each side of the composite capacitor the same parasitic capacitance to ground, and is essential for preserving the symmetry on which the concept of the device relies.

A scanning electron microscope image of a capacitor is shown in Fig.~\ref{fig:CapacitorDesign}b, which shows the Nb strips that form the top plate of one of the $c/2$ parallel plate capacitors.  In the right side of the image, the capacitor is connected to a second parallel plate capacitor (mostly out of view) in the manner described above.

\subsection{Use of normal metal}
\label{app:au}
Superconducting loops in the circulator can trap magnetic flux and lead to an unstable flux environment, interfering with the flux biasing used to control the device. To avoid trapping unwanted flux, we layout the circulator using small amounts of a normal metal (Au).  The thickness (height) of the gold layer is $d = 225$ nm, giving it a sheet resistance of $R_n = 60$ milliohms/square.  To reduce resistive losses in this layer, 11 Au squares are placed in parallel, yielding a total film resistance of about 10 milliohms.  Four of these resistors are placed in each Wheatstone bridge (yellow rectangles in Fig.~\ref{fig:ChipPhoto}c and resistor symbols in Fig.~\ref{fig:BridgeDesign}b) to break supercurrent loops and maintain the symmetry required by the circuit.  Estimates with time-domain numerical simulations (Simulink) indicate that the addition of these resistors causes the dissipation in the circuit to increase by 0.1 dB, limiting the internal $Q$ of the circuit to be less than 2000.

To design the normal metal resistors in a way that prevents proximitization by the nearby niobium, the resistor lengths $l$---defined as its dimension parallel to the flow of current---is constrained to be $l\gg \xi_d$.  Here $\xi_d$ is the coherence length of Au calculated in a dirty limit, where the metal film's mean free path $l_n$ is less than the clean-limit coherence length~\cite{vanduzer:1981}
\begin{equation}
    \xi_c = \frac{\hbar v_F}{2 \pi k_B T}.
\end{equation}
In the above, $v_F$ is the Fermi-velocity of the metal, $k_B$ is Boltzmann's constant, and $T$ is the metal's temperature.  We justify this treatment with the observation that the Au film's mean free path $l_n \approx 600$ nm $\ll 6$ \micron $\approx \xi_c$.  This estimate for $\xi_c$ is made with the assumption that $v_F = 1.4\times10^{6}$ m/s in Au~\cite{ashcroft:1976}, and the temperature $T$ set to 300 mK.  The mean free path is calculated with the Drude model~\cite{ashcroft:1976} and the film's resistivity.

In the dirty-limit, the coherence length $\xi_d$ is essentially a geometric mean of the clean-limit coherence length and the metal's mean free path~\cite{vanduzer:1981}: 
\begin{equation}
\xi_d = \sqrt{\frac{l_n \xi_c}{3}},
\end{equation}  
which comes out to $\xi_d \approx 1$ \micron for the above values of $\xi_c$ and $l_n$.  This value is comparable with measurements of the dirty-limit coherence length in thin films of a similar elemental metal, copper, when one corrects for sample thickness~\cite{pothier:1994,dubos:2001}. 

The condition $l \gg \xi_d$ can be made quantitative with consideration of the superconducting-normal-superconducting (SNS) junction physics which govern the Nb-Au-Nb interface.  Unlike a superconducting-insulator-superconducting junction, which is governed by an energy scale set by the superconducting gap, the natural energy scale for the proximity effect in an SNS junction is the Thouless energy~\cite{dubos:2001}.  For junctions of reasonable size ($l>\xi_d$), the critical current of the SNS junction is
\begin{equation}
I_n = \frac{2 \pi k_b T}{R_n q} \left(\frac{\xi_d}{l}\right)^2 e^{-l/\xi_d} e^{-l/l_{\phi}}.
\end{equation}
Here $q$ is the electron charge, $l_{\phi} \approx 2$ \micron is the inelastic scattering length of gold at 300 mK~\cite{mohanty:2003}, and the Thouless energy is expressed in terms of the coherence length $\xi_d$ and the junction length $l$.  The critical current $I_n$ sets the Josephson energy $E_J = \varphi_0 I_n$ of the SNS junction, and the resistance $R$ of the junction scales in relation to the Josephson energy and the energy in the thermal environment:
\begin{equation}
R = R_n e^{-E_J/k_BT}.
\label{R}
\end{equation}
Along with the Josephson inductance of a SQUID array (Eq.~(\ref{lsquid})), the resistance in Eq.~(\ref{R}) sets an $L/R$ time which characterizes the time required for trapped-flux to dissipate out of the circuit.  Choosing $l > 5.5 \xi_d$ ensures that $L/R$ time is less than 1 second. In preliminary designs, we therefore set $l = 10$ \micron.  In later designs we found experimentally that $l = 5$ \micron also prevents flux-trapping, likely due to a dirty-limit coherence length which is less than our 1 \micron estimate.  The device presented here has $l=5$ \micron.

\subsection{Bias line design}
\label{sec:biaslines}

The circulator's active components are actuated with flux controls created by a pair of on-chip bias lines.  Design of these bias lines involves two important layout considerations: namely, isolating the microwave fields from the bias lines, and preventing the RF bias signals from interfering with the operation of the microwave circuit.  

\begin{figure}[!htb]
\begin{center}
\includegraphics[width=1.0\linewidth]{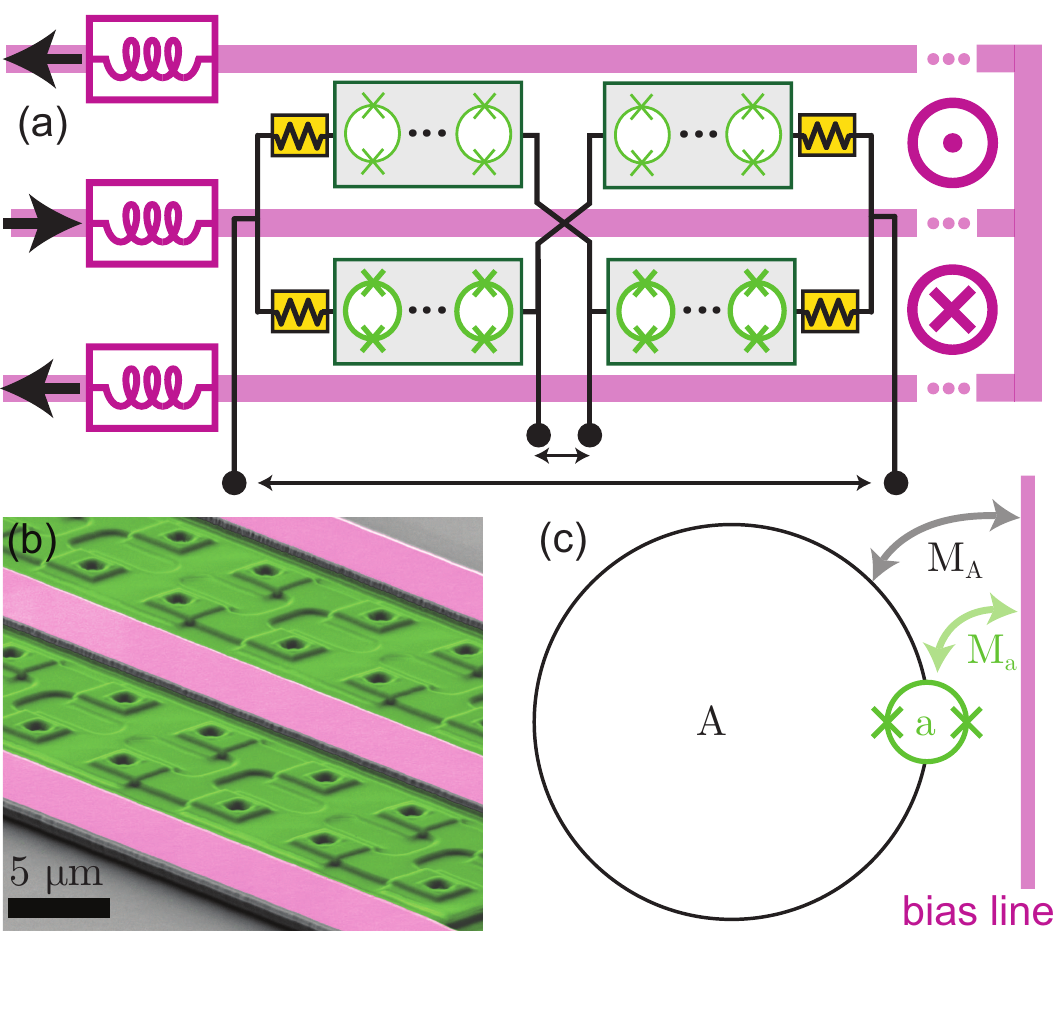}
\caption {
(a) Schematic showing the position of normal metal resistors in the inductive bridges, and the routing of a quadrupole-source bias line which couples strongly to the SQUIDs it encloses and weakly to surrounding loops of the circuit. (b) False color scanning electron microscope image showing two adjacent SQUID arrays (green) and the microstrip lines (pink) that flux-bias them.  (c) Schematic illustrating the challenge of coupling a bias line strongly to a small SQUID loop while coupling it weakly to a larger circuit loop.}
\label{fig:BridgeDesign}
\end{center}
\end{figure}

Isolating the circulator's microwave fields from the bias lines is important because from the perspective of the microwave circuit, coupling to the bias lines acts as an additional loss channel.  To reduce losses of this kind low-pass filters (20 nH spiral inductors) are inserted into the bias lines as they enter and exit the chip (pink inductor symbols in Fig.~\ref{fig:BridgeDesign}a). These simple filters present an impedance of approximately 15 ohms to the bias signals at $\Omega = 2 \pi \times 120$ MHz, whereas at microwave frequencies in the 4 to 8 GHz band their impedance exceeds 500 ohms.  Simulations using commercial planar method-of-moments solvers (AWR Microwave Office) indicate that these filters limit microwave transmission out the bias lines to less than $-20$ dB.

The challenge of the second consideration---preventing bias signals from interfering with the circuit's microwave operation---is illustrated in Fig.~\ref{fig:BridgeDesign}c. The lumped-element representation of the circulator (Fig.~\ref{fig:ChipPhoto}a) contains tunable inductors, realized with flux-modulated SQUIDs, as well as larger circuit loops which are (partly) comprised of SQUIDs.  For simplicity, we consider the effect of a bias line on one such loop of area $A$ which includes a SQUID with area $a$ inside it (Fig.~\ref{fig:BridgeDesign}c).  

To operate the circulator, the bias line must dynamically thread a flux through the SQUID, on the order of a  tenth of a flux quantum $\Phi_g \approx \Phi_0/10$, at a rate $\Omega$.  If the mutual inductance between the bias line and the SQUID loop is denoted as $M_a$, this requires an AC bias current with amplitude $I_g \approx \Phi_0/\left(10 M_a\right)$.  

The time-dependent gradiometric flux, however, also threads through the larger circuit loop of area $A$.  Faraday's law describes the electromotive force induced around this loop, which for a cosinusoidal bias current is $\mathcal{E} = \Omega M_A I_g \sin(\Omega t)$.  We assume the impedance of the loop $Z_A$ is entirely inductive in origin. The loop inductance $L_A$ is the sum of its geometric $L_g$ and Josephson inductance $L_J$, which we write in terms of the participation ratio $p \equiv L_A/L_J$ as $L_A = p L_J$, yielding $Z_A = i \Omega p L_J$.  Ohm's law then allows a calculation of the AC current induced around the loop of area $A$, which with the appropriate substitutions has an amplitude of
\begin{eqnarray}
I_\textrm{ind} \approx \frac{2 \pi}{10 p}  \frac{M_A}{M_a} I_s,
\label{Iind}
\end{eqnarray}
with $I_s$ the critical current of the SQUID.

When the induced currents approach the SQUID critical currents in magnitude, the higher order corrections in Eq.~(\ref{lsquid}) become significant, and when it exceeds the critical currents, the SQUIDs become dissipative elements.  From Eq.~(\ref{Iind}) one can see that the bias signals will couple to the microwave circuit and interfere with the circulator's operation unless the prefactor on the equation's right-hand side is much less than one.  The circulator's performance (in particular, the ability to impedance match the device) requires that the participation ratio $p$ not be much greater than one.  The only way to satisify the requirement, then, is to engineer the mutual inductances such that $\frac{M_A}{M_a} \ll 1$.  This is challenging, as the size of the parallel plate capacitors and the SQUID arrays mandates that $A/a \approx 10^3$.

To overcome the disparity in loop areas and satisfy the coupling condition $M_A/M_a \ll 1$, we layout the bias lines in a symmetric way, such that their currents create magnetic quadrupoles. The layout of the shielded bias lines is shown schematically in Fig.~\ref{fig:BridgeDesign}a, and is also visible in the SEM image in Fig.~\ref{fig:BridgeDesign}b, as well as Fig.~\ref{fig:ChipPhoto}c.  The central bias line, bisecting the bridge, carries the full bias current $I_g$ across the chip, and then splits into two parallel arms, each carrying a current $I_g/2$ on the outside edges of the bridge.  As the currents in these lines flow in opposite directions, the magnetic field $B_g$ from this shielded configuration scales as 
\begin{equation}
B_g = \frac{\mu_0 I_g}{2 \pi r} \left(\frac{\epsilon}{r}\right)^2 + \mathcal{O}\left(\frac{\epsilon}{r}\right)^4,
\end{equation}
where $\epsilon$ is the separation between the inner and outer bias lines, and $\mu_0$ is the vacuum permeability.  We make $\epsilon$ as small as possible in our layouts, given the requirement that the SQUID arrays must reside between the inner and outer bias lines.  These constraints result in the choice $\epsilon = 17.5$ $\mu$m. 

\section{Circulator non-idealities}
\label{app:imperfections}

\begin{figure}[!htb]
\begin{center}
\includegraphics[width=1\linewidth]{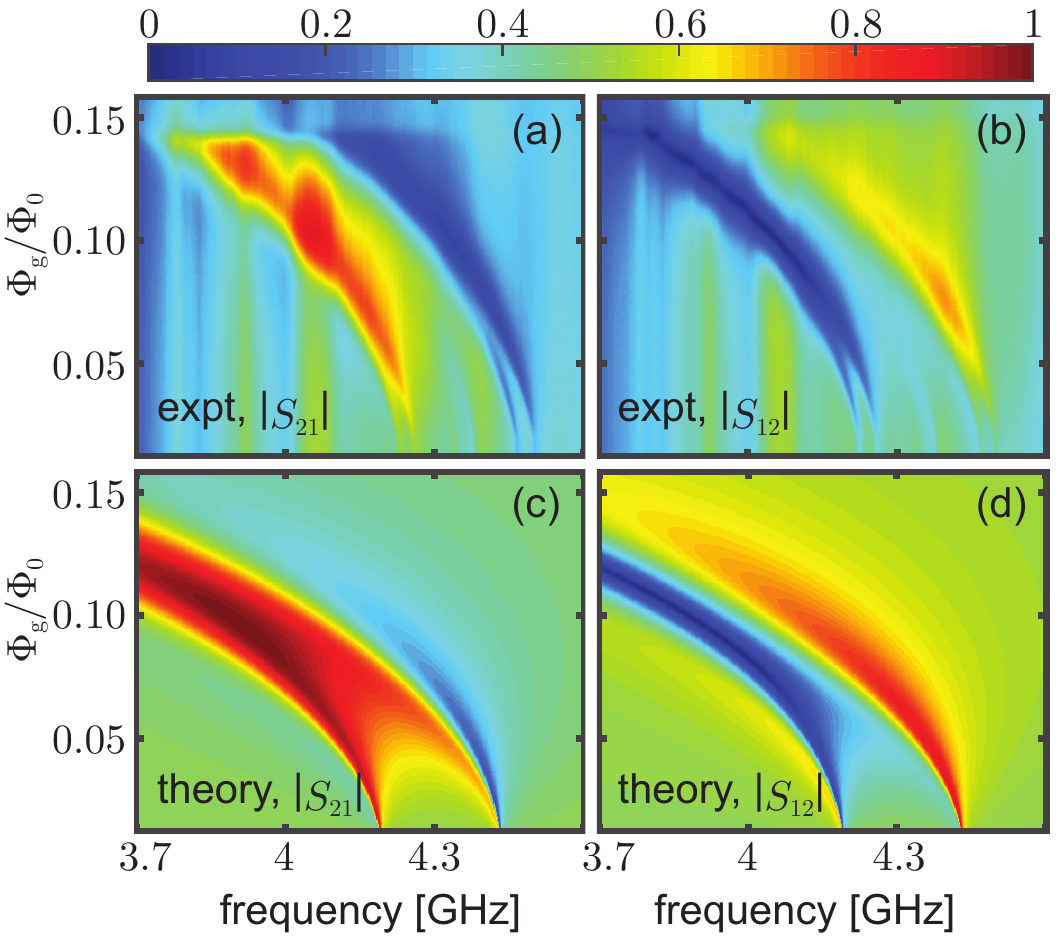}
\caption {
Measurements (a,b) and theoretical predictions (c,d) of $|S_{21}|$ and $|S_{12}|$, as a function of probe frequency and amplitude of the oscillatory gradiometric flux, when the device is configured as a counterclockwise circulator.  Theoretical predictions are made with the expressions in Ref.~\cite{kerckhoff:2015} and the mapping in App.~\ref{app:KL}.  Circuit parameters are fixed at their design targets ($l_0$ = 1 nH, $c$ = 1 pF) and the flux controls are set to match the measurements in (a) and (b): $\Omega = 2 \pi \times 120$ MHz, $\Phi_u = 0.38 \Phi_0$.
}
\label{fig:ampsweep}
\end{center}
\end{figure}

In this section we discuss non-idealities observed in the circulator, in which the network parameter measurements depart from the theoretical predictions of the scattering matrix, obtained with the analytical model in Ref.~\cite{kerckhoff:2015}.  That reference predicts the dependence of $\mathbf{S}$ on the parameters $l_0$, $\delta$, and $\Omega$, and using the relations in App.~\ref{app:KL}, $l_0$ and $\delta$ can be mapped to the flux controls $\Phi_u$ and $\Phi_g$.  To facilitate this comparison, Fig.~\ref{fig:ampsweep} shows measured and predicted transmission parameters, as a function of the probe frequency and the gradiometric flux $\Phi_g$.

Qualitatively, the experiment and model agree fairly well: all four plots show a pair of resonant modes split by twice the modulation rate $\Omega/2 \pi$, in analogy with a Sagnac interferometer~\cite{scully:1997}. As $\Phi_g$ increases, these modes shift down in frequency and broaden.  Furthermore, the device's non-reciprocity is evident in both the model and in experiments: at the lower frequency mode, $S_{21}$ is large in magnitude at the same frequency and gradiometric flux that $S_{12}$ is small.

One can also see aspects of the experimental data which are not captured in the model.  For example, as $\Phi_g$ approaches $0$ the resonant modes become increasingly difficult to perceive in the experimental data. In the model, though, the modes become narrower and $\Phi_g$ decreases, but remain well distinguished from the off-resonant transmission.  This discrepancy is a result of the fact that internal losses are not included in the theoretical model.  In the measurement, the presence of loss means that for small enough $\Phi_g$, the modes become under-coupled and are difficult to detect. 

Another discrepancy is the slight splitting ($2g \approx 2 \times 2 \pi \times 17$ MHz), visible in the experimental plots (Fig.~\ref{fig:ampsweep}a and~b), of each resonant mode.  We attribute this splitting to a hybridization of the circuit's two degenerate resonant modes, which is not included in the model.  


The sharp ``edge'' visible at large $\Phi_g/\Phi_0 \approx 0.15$ in the measurements is an additional difference between the model and experiments. In dedicated studies of this feature, we observe that the location of the edge depends on both the frequency $\Omega$ and the phase $\phi$ between the gradiometric flux drives.  When the inverse of the edge's location is plotted as a function of $\phi$, it scales as $\sin(\phi/2)$, which is precisely the scaling one would expect if the edge resulted from the total flux (i.e. the interference between the two gradiometric flux lines) through a large loop in the circuit exceeding some critical value---for example, a value set by the SQUID critical currents, in the manner discussed in App.~\ref{sec:biaslines}.  This observation supports the conjecture that the edge is caused by induced currents in the microwave circuit which exceed the critical current of the SQUIDs. 

Refinements in the layout can reduce these induced currents, though device operation would still be limited in the amplitude of the applied gradiometric flux; the application of a total external flux $\Phi_u \pm \Phi_g$ with magnitude greater than $\Phi_0/2$ causes a deviation from the simple flux-tunable circuit model described in Fig.~\ref{fig:modprim}.   When the total flux exceeds this threshold, further increase in $\Phi_g$ serves to \emph{balance} the inductive bridges, rather than imbalance them, and a departure from the model is expected in this regime.


A final difference between the model and experiments is visible in the scaling of the resonant modes with the gradiometric flux.  The modes in the theory plots are more sensitive to $\Phi_g$, bending down to lower frequencies than the measured modes.  They also broaden and merge, to a degree which is not apparent in the measurements.  We attribute this discrepancy to geometric inductance in the circuit which reduces the tunability of the resonant delay and restricts the modal linewidth.  

This interpretation is supported by our observation of optimal circulator performance at drive phases $\phi$ distinct from the theoretically expected values at $\pi/2$ and $3 \pi/2$.  When geometric inductance restricts the linewidths of the circulator's resonant modes, it prevents the creation of the brief (2 ns) resonant delay needed to satisfy the convert-delay operation condition: $\Omega \tau = \pi/2$.  The condition can be met with reduction of $\Omega$, but this is undesirable for two reasons.  First, the circulator's bandwidth is proportional to $\Omega$.  
Second, device performance requires that the modulation rate $\Omega$ exceed the internal splitting $g$ of the hybridized resonant modes: $\Omega \gg g$.  

A simple extension of the theory discussed in Sec.~\ref{sec:modelsystem} shows how the circulator's transmission depends on $\phi$ and $\Omega \tau$ in the general case when $\Omega \tau$ takes values other than $\pi/2$:
\begin{eqnarray}
\label{S12}
S_{21} &\propto& \frac{1}{2}\big(1 - \cos\left(\Omega \tau + \phi\right)\big)\\ \nonumber
S_{12} &\propto& \frac{1}{2}\big(1 - \cos(\Omega \tau - \phi)\big).
\end{eqnarray}
From these expressions, it is clear that if $\Omega \tau$ is forced to take values greater than $\pi/2$, improved counterclockwise (clockwise) circulation can be obtained with phases greater (less) than $\pi/2$ ($3\pi/2$).  Our observation of optimal control phases at $\approx 2\pi/3$ and $4 \pi/3$ corresponds to a minimum achievable delay of about 3 ns.

\section{Filtering, attenuation, and power-consumption considerations}
\label{app:filtering}
One of the costs associated with replacing passive ferrite circulators with active on-chip circulators is the power consumption of the control tones, and the heat loads this creates in a dilution refrigerator. Estimating that power consumption requires a discussion of the attenuation and filtering of the control lines.


To determine the attenuation required to keep the added noise below half a photon, the added noise is estimated as a function of the temperature $T$ to which the control lines are thermalized. Scaling and filtering considerations are then discussed, in light of this result.

For simplicity, consider the noise added by the circulator during transmission from its first port to its second port.  Fluctuations of the bias current amplitude and relative phase between the two bias signals will modulate a transmitted tone, thus creating noisy modulation sidebands of the tone. The sideband noise powers caused by amplitude fluctuations $S_I^{\textrm{AN}}$ and phase fluctuations $S_I^{\textrm{PN}}$ are (at most)
\begin{eqnarray}
    S_I^{\textrm{AN}} &=& \left( \frac{\partial S_{21}}{\partial I_g} I_{1\textrm{dB}}\right)^2 S_I, \nonumber \\
    S_I^{\textrm{PN}} &=& \left( \frac{\partial S_{21}}{\partial \phi} \frac{I_{1\textrm{dB}}}{I_g}\right)^2 S_I.
    \label{noisepows}
\end{eqnarray}
Here,  $I_{1\textrm{dB}}$ is the signal current in the device at its $1$ dB compression point and $S_I =  4 k_B T/Z_0$ is the current spectral density of the Johnson noise (in the bias lines) at a temperature $T$~\cite{johnson:1928,nyquist:1928}. Because we operate near a maximum in $|S_{21}|$, the dominant effect of noise in both the amplitude and phase of the bias currents is to modulate the phase of the transmitted tone; i.e., both $S_I^{\textrm{AN}}$ and $S_I^{\textrm{PN}}$ are predominantly phase noise in the transmitted tone.  

The partial derivatives in Eq.~(\ref{noisepows}) can be calculated directly from measurements of the scattering parameters, made as a function of the bias current amplitude $I_g$ and the phase between the bias lines $\phi$ (shown, for example, in Fig.~\ref{fig:ampsweep} and Fig.~\ref{fig:phasesweep}).  After these numerical derivatives are calculated, the sideband noise powers may be divided by $2 \hbar \omega_p / Z_0$ to convert them to photon numbers. In our measurements, where the bias lines are thermalized to $T \approx 300$ K, this results in $n = 14$ photons of added noise, with $S_I^{\textrm{AN}}$ accounting for $95\%$ of the noise.  

Positioning $40$ db of attenuation at room temperature and $20$ dB at the four Kelvin stage of a dilution refrigerator would result in a noise temperature of $T \approx 7$ K, or in units of photons, $n = 0.3$.  This level of attenuation is reasonable for modern dilution refrigerators, as the circulator operates with gradiometric currents $I_g$ on the scale of $100$ $\mu$A: the heat load caused by a $20$ dB attenuator at the four-K stage is $50$ $\mu$W, which is much less than the Watt-scale cooling power available at that stage.  

With superconducting twisted-pairs to carry the bias currents from the four-K stage to the mixing chamber plate, and a contact resistance of $10$ milliohms at the chip interface, the heat load on the mixing chamber plate is $100$ pW. This load is also much less than the roughly $50$ $\mu$W of available cooling power on a $50$ mK mixing chamber plate. These considerations are summarized in Tab.~(\ref{tab:powertab}), which presents a power budget for an active circulator with control lines thermalized as described above.

\begin{table}[!htb]
\centering
\caption{Power budget for an active circulator with control lines thermalized to $T \approx 7$ K as described in the text. In this configuration, the circulator adds $0.3$ photons of noise.   A contact resistance of $10$ milliohms is assumed at the chip interface.  Cited cooling powers are for an Oxford cryogen-free dilution refrigerator, and are approximate.}
\label{tab:powertab}
\begin{tabular}{|l|l|l|l|}
\hline
$T$ [K]       & $I_g$ [A]    & heat load [W]    & cooling power [W] \\ \hline
300 & $10^{-1}$  & $5 \times 10^{-1}$ & n/a \\ \hline
4 & $10^{-3}$  & $5 \times 10^{-5}$ & $7.5 \times 10^{-1}$ \\ \hline
0.05 & $10^{-4}$  & $10^{-10}$ & $5 \times 10^{-5}$ \\ \hline
\end{tabular}
\end{table}

This analysis indicates the feasibility of operating $10^3$ on-chip circulators in a single dilution refrigerator, each with less than half a photon of added noise.  We emphasize that it is one of many possible design choices and it is possible to reduce the added noise and dissipated power in several different ways. For example, the bias lines could be filtered to reject the noise below 50 MHz, which adds noise in the circulator's band, while still passing 100 MHz bias tones.         



\begin{thebibliography}{53}%
\makeatletter
\providecommand \@ifxundefined [1]{%
 \@ifx{#1\undefined}
}%
\providecommand \@ifnum [1]{%
 \ifnum #1\expandafter \@firstoftwo
 \else \expandafter \@secondoftwo
 \fi
}%
\providecommand \@ifx [1]{%
 \ifx #1\expandafter \@firstoftwo
 \else \expandafter \@secondoftwo
 \fi
}%
\providecommand \natexlab [1]{#1}%
\providecommand \enquote  [1]{``#1''}%
\providecommand \bibnamefont  [1]{#1}%
\providecommand \bibfnamefont [1]{#1}%
\providecommand \citenamefont [1]{#1}%
\providecommand \href@noop [0]{\@secondoftwo}%
\providecommand \href [0]{\begingroup \@sanitize@url \@href}%
\providecommand \@href[1]{\@@startlink{#1}\@@href}%
\providecommand \@@href[1]{\endgroup#1\@@endlink}%
\providecommand \@sanitize@url [0]{\catcode `\\12\catcode `\$12\catcode
  `\&12\catcode `\#12\catcode `\^12\catcode `\_12\catcode `\%12\relax}%
\providecommand \@@startlink[1]{}%
\providecommand \@@endlink[0]{}%
\providecommand \url  [0]{\begingroup\@sanitize@url \@url }%
\providecommand \@url [1]{\endgroup\@href {#1}{\urlprefix }}%
\providecommand \urlprefix  [0]{URL }%
\providecommand \Eprint [0]{\href }%
\providecommand \doibase [0]{http://dx.doi.org/}%
\providecommand \selectlanguage [0]{\@gobble}%
\providecommand \bibinfo  [0]{\@secondoftwo}%
\providecommand \bibfield  [0]{\@secondoftwo}%
\providecommand \translation [1]{[#1]}%
\providecommand \BibitemOpen [0]{}%
\providecommand \bibitemStop [0]{}%
\providecommand \bibitemNoStop [0]{.\EOS\space}%
\providecommand \EOS [0]{\spacefactor3000\relax}%
\providecommand \BibitemShut  [1]{\csname bibitem#1\endcsname}%
\let\auto@bib@innerbib\@empty
\bibitem [{\citenamefont {Blais}\ \emph {et~al.}(2004)\citenamefont {Blais},
  \citenamefont {Huang}, \citenamefont {Wallraff}, \citenamefont {Girvin},\
  and\ \citenamefont {Schoelkopf}}]{blais:2004}%
  \BibitemOpen
  \bibfield  {author} {\bibinfo {author} {\bibfnamefont {A.}~\bibnamefont
  {Blais}}, \bibinfo {author} {\bibfnamefont {R.-S.}\ \bibnamefont {Huang}},
  \bibinfo {author} {\bibfnamefont {A.}~\bibnamefont {Wallraff}}, \bibinfo
  {author} {\bibfnamefont {S.~M.}\ \bibnamefont {Girvin}}, \ and\ \bibinfo
  {author} {\bibfnamefont {R.~J.}\ \bibnamefont {Schoelkopf}},\ }\href@noop {}
  {\bibfield  {journal} {\bibinfo  {journal} {Physical Review A}\ }\textbf
  {\bibinfo {volume} {69}},\ \bibinfo {pages} {062320} (\bibinfo {year}
  {2004})}\BibitemShut {NoStop}%
\bibitem [{\citenamefont {Kelly}\ \emph {et~al.}(2015)\citenamefont {Kelly},
  \citenamefont {Barends}, \citenamefont {Fowler}, \citenamefont {Megrant},
  \citenamefont {Jeffrey}, \citenamefont {White}, \citenamefont {Sank},
  \citenamefont {Mutus}, \citenamefont {Campbell}, \citenamefont {Chen} \emph
  {et~al.}}]{kelly:2015}%
  \BibitemOpen
  \bibfield  {author} {\bibinfo {author} {\bibfnamefont {J.}~\bibnamefont
  {Kelly}}, \bibinfo {author} {\bibfnamefont {R.}~\bibnamefont {Barends}},
  \bibinfo {author} {\bibfnamefont {A.~G.}\ \bibnamefont {Fowler}}, \bibinfo
  {author} {\bibfnamefont {A.}~\bibnamefont {Megrant}}, \bibinfo {author}
  {\bibfnamefont {E.}~\bibnamefont {Jeffrey}}, \bibinfo {author} {\bibfnamefont
  {T.~C.}\ \bibnamefont {White}}, \bibinfo {author} {\bibfnamefont
  {D.}~\bibnamefont {Sank}}, \bibinfo {author} {\bibfnamefont {J.~Y.}\
  \bibnamefont {Mutus}}, \bibinfo {author} {\bibfnamefont {B.}~\bibnamefont
  {Campbell}}, \bibinfo {author} {\bibfnamefont {Y.}~\bibnamefont {Chen}},
  \emph {et~al.},\ }\href@noop {} {\bibfield  {journal} {\bibinfo  {journal}
  {Nature}\ }\textbf {\bibinfo {volume} {519}},\ \bibinfo {pages} {66}
  (\bibinfo {year} {2015})}\BibitemShut {NoStop}%
\bibitem [{\citenamefont {Ofek}\ \emph {et~al.}(2016)\citenamefont {Ofek},
  \citenamefont {Petrenko}, \citenamefont {Heeres}, \citenamefont {Reinhold},
  \citenamefont {Leghtas}, \citenamefont {Vlastakis}, \citenamefont {Liu},
  \citenamefont {Frunzio}, \citenamefont {Girvin}, \citenamefont {Jiang},
  \citenamefont {Mirrahimi}, \citenamefont {Devoret},\ and\ \citenamefont
  {Schoelkopf}}]{ofek:2016}%
  \BibitemOpen
  \bibfield  {author} {\bibinfo {author} {\bibfnamefont {N.}~\bibnamefont
  {Ofek}}, \bibinfo {author} {\bibfnamefont {A.}~\bibnamefont {Petrenko}},
  \bibinfo {author} {\bibfnamefont {R.}~\bibnamefont {Heeres}}, \bibinfo
  {author} {\bibfnamefont {P.}~\bibnamefont {Reinhold}}, \bibinfo {author}
  {\bibfnamefont {Z.}~\bibnamefont {Leghtas}}, \bibinfo {author} {\bibfnamefont
  {B.}~\bibnamefont {Vlastakis}}, \bibinfo {author} {\bibfnamefont
  {Y.}~\bibnamefont {Liu}}, \bibinfo {author} {\bibfnamefont {L.}~\bibnamefont
  {Frunzio}}, \bibinfo {author} {\bibfnamefont {S.}~\bibnamefont {Girvin}},
  \bibinfo {author} {\bibfnamefont {L.}~\bibnamefont {Jiang}}, \bibinfo
  {author} {\bibfnamefont {M.}~\bibnamefont {Mirrahimi}}, \bibinfo {author}
  {\bibfnamefont {M.~H.}\ \bibnamefont {Devoret}}, \ and\ \bibinfo {author}
  {\bibfnamefont {R.~J.}\ \bibnamefont {Schoelkopf}},\ }\href@noop {}
  {\bibfield  {journal} {\bibinfo  {journal} {Nature}\ } (\bibinfo {year}
  {2016})}\BibitemShut {NoStop}%
\bibitem [{\citenamefont {Castellanos-Beltran}\ \emph
  {et~al.}(2008)\citenamefont {Castellanos-Beltran}, \citenamefont {Irwin},
  \citenamefont {Hilton}, \citenamefont {Vale},\ and\ \citenamefont
  {Lehnert}}]{castellanos:2008}%
  \BibitemOpen
  \bibfield  {author} {\bibinfo {author} {\bibfnamefont {M.~A.}\ \bibnamefont
  {Castellanos-Beltran}}, \bibinfo {author} {\bibfnamefont {K.~D.}\
  \bibnamefont {Irwin}}, \bibinfo {author} {\bibfnamefont {G.~C.}\ \bibnamefont
  {Hilton}}, \bibinfo {author} {\bibfnamefont {L.~R.}\ \bibnamefont {Vale}}, \
  and\ \bibinfo {author} {\bibfnamefont {K.~W.}\ \bibnamefont {Lehnert}},\
  }\href@noop {} {\bibfield  {journal} {\bibinfo  {journal} {Nature Physics}\
  }\textbf {\bibinfo {volume} {4}},\ \bibinfo {pages} {929} (\bibinfo {year}
  {2008})}\BibitemShut {NoStop}%
\bibitem [{\citenamefont {Bergeal}\ \emph {et~al.}(2010)\citenamefont
  {Bergeal}, \citenamefont {Vijay}, \citenamefont {Manucharyan}, \citenamefont
  {Siddiqi}, \citenamefont {Schoelkopf}, \citenamefont {Girvin},\ and\
  \citenamefont {Devoret}}]{bergeal:2010}%
  \BibitemOpen
  \bibfield  {author} {\bibinfo {author} {\bibfnamefont {N.}~\bibnamefont
  {Bergeal}}, \bibinfo {author} {\bibfnamefont {R.}~\bibnamefont {Vijay}},
  \bibinfo {author} {\bibfnamefont {V.~E.}\ \bibnamefont {Manucharyan}},
  \bibinfo {author} {\bibfnamefont {I.}~\bibnamefont {Siddiqi}}, \bibinfo
  {author} {\bibfnamefont {R.~J.}\ \bibnamefont {Schoelkopf}}, \bibinfo
  {author} {\bibfnamefont {S.~M.}\ \bibnamefont {Girvin}}, \ and\ \bibinfo
  {author} {\bibfnamefont {M.~H.}\ \bibnamefont {Devoret}},\ }\href@noop {}
  {\bibfield  {journal} {\bibinfo  {journal} {Nature Physics}\ }\textbf
  {\bibinfo {volume} {6}},\ \bibinfo {pages} {296} (\bibinfo {year}
  {2010})}\BibitemShut {NoStop}%
\bibitem [{\citenamefont {Macklin}\ \emph {et~al.}(2015)\citenamefont
  {Macklin}, \citenamefont {O`Brien}, \citenamefont {Hover}, \citenamefont
  {Schwartz}, \citenamefont {Bolkhovsky}, \citenamefont {Zhang}, \citenamefont
  {Oliver},\ and\ \citenamefont {Siddiqi}}]{macklin:2015}%
  \BibitemOpen
  \bibfield  {author} {\bibinfo {author} {\bibfnamefont {C.}~\bibnamefont
  {Macklin}}, \bibinfo {author} {\bibfnamefont {K.}~\bibnamefont {O`Brien}},
  \bibinfo {author} {\bibfnamefont {D.}~\bibnamefont {Hover}}, \bibinfo
  {author} {\bibfnamefont {M.~E.}\ \bibnamefont {Schwartz}}, \bibinfo {author}
  {\bibfnamefont {V.}~\bibnamefont {Bolkhovsky}}, \bibinfo {author}
  {\bibfnamefont {X.}~\bibnamefont {Zhang}}, \bibinfo {author} {\bibfnamefont
  {W.~D.}\ \bibnamefont {Oliver}}, \ and\ \bibinfo {author} {\bibfnamefont
  {I.}~\bibnamefont {Siddiqi}},\ }\href@noop {} {\bibfield  {journal} {\bibinfo
   {journal} {Science}\ }\textbf {\bibinfo {volume} {350}},\ \bibinfo {pages}
  {307} (\bibinfo {year} {2015})}\BibitemShut {NoStop}%
\bibitem [{\citenamefont {Vijay}\ \emph {et~al.}(2011)\citenamefont {Vijay},
  \citenamefont {Slichter},\ and\ \citenamefont {Siddiqi}}]{vijay:2011}%
  \BibitemOpen
  \bibfield  {author} {\bibinfo {author} {\bibfnamefont {R.}~\bibnamefont
  {Vijay}}, \bibinfo {author} {\bibfnamefont {D.~H.}\ \bibnamefont {Slichter}},
  \ and\ \bibinfo {author} {\bibfnamefont {I.}~\bibnamefont {Siddiqi}},\ }\href
  {\doibase 10.1103/PhysRevLett.106.110502} {\bibfield  {journal} {\bibinfo
  {journal} {Phys. Rev. Lett.}\ }\textbf {\bibinfo {volume} {106}},\ \bibinfo
  {pages} {110502} (\bibinfo {year} {2011})}\BibitemShut {NoStop}%
\bibitem [{\citenamefont {Rist\`e}\ \emph {et~al.}(2012)\citenamefont
  {Rist\`e}, \citenamefont {van Leeuwen}, \citenamefont {Ku}, \citenamefont
  {Lehnert},\ and\ \citenamefont {DiCarlo}}]{riste:2012}%
  \BibitemOpen
  \bibfield  {author} {\bibinfo {author} {\bibfnamefont {D.}~\bibnamefont
  {Rist\`e}}, \bibinfo {author} {\bibfnamefont {J.~G.}\ \bibnamefont {van
  Leeuwen}}, \bibinfo {author} {\bibfnamefont {H.-S.}\ \bibnamefont {Ku}},
  \bibinfo {author} {\bibfnamefont {K.~W.}\ \bibnamefont {Lehnert}}, \ and\
  \bibinfo {author} {\bibfnamefont {L.}~\bibnamefont {DiCarlo}},\ }\href
  {\doibase 10.1103/PhysRevLett.109.050507} {\bibfield  {journal} {\bibinfo
  {journal} {Phys. Rev. Lett.}\ }\textbf {\bibinfo {volume} {109}},\ \bibinfo
  {pages} {050507} (\bibinfo {year} {2012})}\BibitemShut {NoStop}%
\bibitem [{\citenamefont {Caves}(1982)}]{caves:1982}%
  \BibitemOpen
  \bibfield  {author} {\bibinfo {author} {\bibfnamefont {C.~M.}\ \bibnamefont
  {Caves}},\ }\href {\doibase 10.1103/PhysRevD.26.1817} {\bibfield  {journal}
  {\bibinfo  {journal} {Phys. Rev. D}\ }\textbf {\bibinfo {volume} {26}},\
  \bibinfo {pages} {1817} (\bibinfo {year} {1982})}\BibitemShut {NoStop}%
\bibitem [{\citenamefont {Pozar}(2012)}]{pozar:2011}%
  \BibitemOpen
  \bibfield  {author} {\bibinfo {author} {\bibfnamefont {D.~M.}\ \bibnamefont
  {Pozar}},\ }\href@noop {} {\enquote {\bibinfo {title} {Microwave engineering.
  4th},}\ } (\bibinfo {year} {2012})\BibitemShut {NoStop}%
\bibitem [{\citenamefont {Fay}\ and\ \citenamefont
  {Comstock}(1965)}]{fay:1965}%
  \BibitemOpen
  \bibfield  {author} {\bibinfo {author} {\bibfnamefont {C.~E.}\ \bibnamefont
  {Fay}}\ and\ \bibinfo {author} {\bibfnamefont {R.~L.}\ \bibnamefont
  {Comstock}},\ }\href@noop {} {\bibfield  {journal} {\bibinfo  {journal}
  {Microwave Theory and Techniques, IEEE Transactions on}\ }\textbf {\bibinfo
  {volume} {13}},\ \bibinfo {pages} {15} (\bibinfo {year} {1965})}\BibitemShut
  {NoStop}%
\bibitem [{\citenamefont {Viola}\ and\ \citenamefont
  {DiVincenzo}(2014)}]{viola:2014}%
  \BibitemOpen
  \bibfield  {author} {\bibinfo {author} {\bibfnamefont {G.}~\bibnamefont
  {Viola}}\ and\ \bibinfo {author} {\bibfnamefont {D.~P.}\ \bibnamefont
  {DiVincenzo}},\ }\href@noop {} {\bibfield  {journal} {\bibinfo  {journal}
  {Physical Review X}\ }\textbf {\bibinfo {volume} {4}},\ \bibinfo {pages}
  {021019} (\bibinfo {year} {2014})}\BibitemShut {NoStop}%
\bibitem [{\citenamefont {Mahoney}\ \emph
  {et~al.}(2017{\natexlab{a}})\citenamefont {Mahoney}, \citenamefont {Colless},
  \citenamefont {Pauka}, \citenamefont {Hornibrook}, \citenamefont {Watson},
  \citenamefont {Gardner}, \citenamefont {Manfra}, \citenamefont {Doherty},\
  and\ \citenamefont {Reilly}}]{mahoney:2017}%
  \BibitemOpen
  \bibfield  {author} {\bibinfo {author} {\bibfnamefont {A.~C.}\ \bibnamefont
  {Mahoney}}, \bibinfo {author} {\bibfnamefont {J.~I.}\ \bibnamefont
  {Colless}}, \bibinfo {author} {\bibfnamefont {S.~J.}\ \bibnamefont {Pauka}},
  \bibinfo {author} {\bibfnamefont {J.~M.}\ \bibnamefont {Hornibrook}},
  \bibinfo {author} {\bibfnamefont {J.~D.}\ \bibnamefont {Watson}}, \bibinfo
  {author} {\bibfnamefont {G.~C.}\ \bibnamefont {Gardner}}, \bibinfo {author}
  {\bibfnamefont {M.~J.}\ \bibnamefont {Manfra}}, \bibinfo {author}
  {\bibfnamefont {A.~C.}\ \bibnamefont {Doherty}}, \ and\ \bibinfo {author}
  {\bibfnamefont {D.~J.}\ \bibnamefont {Reilly}},\ }\href@noop {} {\bibfield
  {journal} {\bibinfo  {journal} {Phys. Rev. X}\ }\textbf {\bibinfo {volume}
  {7}},\ \bibinfo {pages} {011007} (\bibinfo {year}
  {2017}{\natexlab{a}})}\BibitemShut {NoStop}%
\bibitem [{\citenamefont {Mahoney}\ \emph
  {et~al.}(2017{\natexlab{b}})\citenamefont {Mahoney}, \citenamefont {Colless},
  \citenamefont {Peeters}, \citenamefont {Pauka}, \citenamefont {Fox},
  \citenamefont {Kou}, \citenamefont {Pan}, \citenamefont {K.~L.~Wang},\ and\
  \citenamefont {Reilly}}]{mahoney:2017b}%
  \BibitemOpen
  \bibfield  {author} {\bibinfo {author} {\bibfnamefont {A.~C.}\ \bibnamefont
  {Mahoney}}, \bibinfo {author} {\bibfnamefont {J.~I.}\ \bibnamefont
  {Colless}}, \bibinfo {author} {\bibfnamefont {L.}~\bibnamefont {Peeters}},
  \bibinfo {author} {\bibfnamefont {S.~J.}\ \bibnamefont {Pauka}}, \bibinfo
  {author} {\bibfnamefont {E.~J.}\ \bibnamefont {Fox}}, \bibinfo {author}
  {\bibfnamefont {X.}~\bibnamefont {Kou}}, \bibinfo {author} {\bibfnamefont
  {L.}~\bibnamefont {Pan}}, \bibinfo {author} {\bibfnamefont {D.~G.-G.}\
  \bibnamefont {K.~L.~Wang}}, \ and\ \bibinfo {author} {\bibfnamefont {D.~J.}\
  \bibnamefont {Reilly}},\ }\href@noop {} {\bibfield  {journal} {\bibinfo
  {journal} {arXiv preprint arXiv:1703.03122}\ } (\bibinfo {year}
  {2017}{\natexlab{b}})}\BibitemShut {NoStop}%
\bibitem [{\citenamefont {Anderson}\ and\ \citenamefont
  {Newcomb}(1965)}]{anderson:1965}%
  \BibitemOpen
  \bibfield  {author} {\bibinfo {author} {\bibfnamefont {B.~D.~O.}\
  \bibnamefont {Anderson}}\ and\ \bibinfo {author} {\bibfnamefont {R.~W.}\
  \bibnamefont {Newcomb}},\ }\href@noop {} {\bibfield  {journal} {\bibinfo
  {journal} {Proceedings of the IEEE}\ }\textbf {\bibinfo {volume} {53}},\
  \bibinfo {pages} {1674} (\bibinfo {year} {1965})}\BibitemShut {NoStop}%
\bibitem [{\citenamefont {Anderson}\ and\ \citenamefont
  {Newcomb}(1966)}]{anderson:1966}%
  \BibitemOpen
  \bibfield  {author} {\bibinfo {author} {\bibfnamefont {B.}~\bibnamefont
  {Anderson}}\ and\ \bibinfo {author} {\bibfnamefont {R.}~\bibnamefont
  {Newcomb}},\ }\href@noop {} {\bibfield  {journal} {\bibinfo  {journal}
  {Circuit Theory, IEEE Transactions on}\ }\textbf {\bibinfo {volume} {13}},\
  \bibinfo {pages} {233} (\bibinfo {year} {1966})}\BibitemShut {NoStop}%
\bibitem [{\citenamefont {Kamal}\ \emph {et~al.}(2011)\citenamefont {Kamal},
  \citenamefont {Clarke},\ and\ \citenamefont {Devoret}}]{kamal:2011}%
  \BibitemOpen
  \bibfield  {author} {\bibinfo {author} {\bibfnamefont {A.}~\bibnamefont
  {Kamal}}, \bibinfo {author} {\bibfnamefont {J.}~\bibnamefont {Clarke}}, \
  and\ \bibinfo {author} {\bibfnamefont {M.~H.}\ \bibnamefont {Devoret}},\
  }\href@noop {} {\bibfield  {journal} {\bibinfo  {journal} {Nature Physics}\
  }\textbf {\bibinfo {volume} {7}},\ \bibinfo {pages} {311} (\bibinfo {year}
  {2011})}\BibitemShut {NoStop}%
\bibitem [{\citenamefont {Metelmann}\ and\ \citenamefont
  {Clerk}(2015)}]{metelmann:2015}%
  \BibitemOpen
  \bibfield  {author} {\bibinfo {author} {\bibfnamefont {A.}~\bibnamefont
  {Metelmann}}\ and\ \bibinfo {author} {\bibfnamefont {A.}~\bibnamefont
  {Clerk}},\ }\href@noop {} {\bibfield  {journal} {\bibinfo  {journal}
  {Physical Review X}\ }\textbf {\bibinfo {volume} {5}},\ \bibinfo {pages}
  {021025} (\bibinfo {year} {2015})}\BibitemShut {NoStop}%
\bibitem [{\citenamefont {Kerckhoff}\ \emph {et~al.}(2015)\citenamefont
  {Kerckhoff}, \citenamefont {Lalumi\`ere}, \citenamefont {Chapman},
  \citenamefont {Blais},\ and\ \citenamefont {Lehnert}}]{kerckhoff:2015}%
  \BibitemOpen
  \bibfield  {author} {\bibinfo {author} {\bibfnamefont {J.}~\bibnamefont
  {Kerckhoff}}, \bibinfo {author} {\bibfnamefont {K.}~\bibnamefont
  {Lalumi\`ere}}, \bibinfo {author} {\bibfnamefont {B.~J.}\ \bibnamefont
  {Chapman}}, \bibinfo {author} {\bibfnamefont {A.}~\bibnamefont {Blais}}, \
  and\ \bibinfo {author} {\bibfnamefont {K.~W.}\ \bibnamefont {Lehnert}},\
  }\href {\doibase 10.1103/PhysRevApplied.4.034002} {\bibfield  {journal}
  {\bibinfo  {journal} {Phys. Rev. Applied}\ }\textbf {\bibinfo {volume} {4}},\
  \bibinfo {pages} {034002} (\bibinfo {year} {2015})}\BibitemShut {NoStop}%
\bibitem [{\citenamefont {Abdo}\ \emph {et~al.}(2014)\citenamefont {Abdo},
  \citenamefont {Sliwa}, \citenamefont {Shankar}, \citenamefont {Hatridge},
  \citenamefont {Frunzio}, \citenamefont {Schoelkopf},\ and\ \citenamefont
  {Devoret}}]{abdo:2014}%
  \BibitemOpen
  \bibfield  {author} {\bibinfo {author} {\bibfnamefont {B.}~\bibnamefont
  {Abdo}}, \bibinfo {author} {\bibfnamefont {K.}~\bibnamefont {Sliwa}},
  \bibinfo {author} {\bibfnamefont {S.}~\bibnamefont {Shankar}}, \bibinfo
  {author} {\bibfnamefont {M.}~\bibnamefont {Hatridge}}, \bibinfo {author}
  {\bibfnamefont {L.}~\bibnamefont {Frunzio}}, \bibinfo {author} {\bibfnamefont
  {R.~J.}\ \bibnamefont {Schoelkopf}}, \ and\ \bibinfo {author} {\bibfnamefont
  {M.~H.}\ \bibnamefont {Devoret}},\ }\href@noop {} {\bibfield  {journal}
  {\bibinfo  {journal} {Physical review letters}\ }\textbf {\bibinfo {volume}
  {112}},\ \bibinfo {pages} {167701} (\bibinfo {year} {2014})}\BibitemShut
  {NoStop}%
\bibitem [{\citenamefont {Estep}\ \emph {et~al.}(2014)\citenamefont {Estep},
  \citenamefont {Sounas}, \citenamefont {Soric},\ and\ \citenamefont
  {Al{\`u}}}]{estep:2014}%
  \BibitemOpen
  \bibfield  {author} {\bibinfo {author} {\bibfnamefont {N.~A.}\ \bibnamefont
  {Estep}}, \bibinfo {author} {\bibfnamefont {D.~L.}\ \bibnamefont {Sounas}},
  \bibinfo {author} {\bibfnamefont {J.}~\bibnamefont {Soric}}, \ and\ \bibinfo
  {author} {\bibfnamefont {A.}~\bibnamefont {Al{\`u}}},\ }\href@noop {}
  {\bibfield  {journal} {\bibinfo  {journal} {Nature Physics}\ }\textbf
  {\bibinfo {volume} {10}},\ \bibinfo {pages} {923} (\bibinfo {year}
  {2014})}\BibitemShut {NoStop}%
\bibitem [{\citenamefont {Ranzani}\ and\ \citenamefont
  {Aumentado}(2015)}]{ranzani:2015}%
  \BibitemOpen
  \bibfield  {author} {\bibinfo {author} {\bibfnamefont {L.}~\bibnamefont
  {Ranzani}}\ and\ \bibinfo {author} {\bibfnamefont {J.}~\bibnamefont
  {Aumentado}},\ }\href@noop {} {\bibfield  {journal} {\bibinfo  {journal} {New
  Journal of Physics}\ }\textbf {\bibinfo {volume} {17}},\ \bibinfo {pages}
  {023024} (\bibinfo {year} {2015})}\BibitemShut {NoStop}%
\bibitem [{\citenamefont {Sliwa}\ \emph {et~al.}(2015)\citenamefont {Sliwa},
  \citenamefont {Hatridge}, \citenamefont {Narla}, \citenamefont {Shankar},
  \citenamefont {Frunzio}, \citenamefont {Schoelkopf},\ and\ \citenamefont
  {Devoret}}]{sliwa:2015}%
  \BibitemOpen
  \bibfield  {author} {\bibinfo {author} {\bibfnamefont {K.~M.}\ \bibnamefont
  {Sliwa}}, \bibinfo {author} {\bibfnamefont {M.}~\bibnamefont {Hatridge}},
  \bibinfo {author} {\bibfnamefont {A.}~\bibnamefont {Narla}}, \bibinfo
  {author} {\bibfnamefont {S.}~\bibnamefont {Shankar}}, \bibinfo {author}
  {\bibfnamefont {L.}~\bibnamefont {Frunzio}}, \bibinfo {author} {\bibfnamefont
  {R.~J.}\ \bibnamefont {Schoelkopf}}, \ and\ \bibinfo {author} {\bibfnamefont
  {M.~H.}\ \bibnamefont {Devoret}},\ }\href@noop {} {\bibfield  {journal}
  {\bibinfo  {journal} {Phys. Rev. X}\ }\textbf {\bibinfo {volume} {5}},\
  \bibinfo {pages} {041020} (\bibinfo {year} {2015})}\BibitemShut {NoStop}%
\bibitem [{\citenamefont {Lecocq}\ \emph {et~al.}(2017)\citenamefont {Lecocq},
  \citenamefont {Ranzani}, \citenamefont {Peterson}, \citenamefont {Cicak},
  \citenamefont {Simmonds}, \citenamefont {Teufel},\ and\ \citenamefont
  {Aumentado}}]{lecocq:2017}%
  \BibitemOpen
  \bibfield  {author} {\bibinfo {author} {\bibfnamefont {F.}~\bibnamefont
  {Lecocq}}, \bibinfo {author} {\bibfnamefont {L.}~\bibnamefont {Ranzani}},
  \bibinfo {author} {\bibfnamefont {G.~A.}\ \bibnamefont {Peterson}}, \bibinfo
  {author} {\bibfnamefont {K.}~\bibnamefont {Cicak}}, \bibinfo {author}
  {\bibfnamefont {R.~W.}\ \bibnamefont {Simmonds}}, \bibinfo {author}
  {\bibfnamefont {J.~D.}\ \bibnamefont {Teufel}}, \ and\ \bibinfo {author}
  {\bibfnamefont {J.}~\bibnamefont {Aumentado}},\ }\href {\doibase
  10.1103/PhysRevApplied.7.024028} {\bibfield  {journal} {\bibinfo  {journal}
  {Phys. Rev. Applied}\ }\textbf {\bibinfo {volume} {7}},\ \bibinfo {pages}
  {024028} (\bibinfo {year} {2017})}\BibitemShut {NoStop}%
\bibitem [{\citenamefont {Abdo}\ \emph {et~al.}(2017)\citenamefont {Abdo},
  \citenamefont {Brink},\ and\ \citenamefont {Chow}}]{abdo:2017}%
  \BibitemOpen
  \bibfield  {author} {\bibinfo {author} {\bibfnamefont {B.}~\bibnamefont
  {Abdo}}, \bibinfo {author} {\bibfnamefont {M.}~\bibnamefont {Brink}}, \ and\
  \bibinfo {author} {\bibfnamefont {J.~M.}\ \bibnamefont {Chow}},\ }\href
  {\doibase 10.1103/PhysRevApplied.8.034009} {\bibfield  {journal} {\bibinfo
  {journal} {Phys. Rev. Applied}\ }\textbf {\bibinfo {volume} {8}},\ \bibinfo
  {pages} {034009} (\bibinfo {year} {2017})}\BibitemShut {NoStop}%
\bibitem [{\citenamefont {Bernier}\ \emph {et~al.}(2017)\citenamefont
  {Bernier}, \citenamefont {Toth}, \citenamefont {Koottandavida}, \citenamefont
  {Ioannou}, \citenamefont {Malz}, \citenamefont {Nunnenkamp}, \citenamefont
  {Feofanov},\ and\ \citenamefont {Kippenberg}}]{bernier:2017}%
  \BibitemOpen
  \bibfield  {author} {\bibinfo {author} {\bibfnamefont {N.~R.}\ \bibnamefont
  {Bernier}}, \bibinfo {author} {\bibfnamefont {L.~D.}\ \bibnamefont {Toth}},
  \bibinfo {author} {\bibfnamefont {A.}~\bibnamefont {Koottandavida}}, \bibinfo
  {author} {\bibfnamefont {M.~A.}\ \bibnamefont {Ioannou}}, \bibinfo {author}
  {\bibfnamefont {D.}~\bibnamefont {Malz}}, \bibinfo {author} {\bibfnamefont
  {A.}~\bibnamefont {Nunnenkamp}}, \bibinfo {author} {\bibfnamefont
  {A.}~\bibnamefont {Feofanov}}, \ and\ \bibinfo {author} {\bibfnamefont
  {T.}~\bibnamefont {Kippenberg}},\ }\href@noop {} {\bibfield  {journal}
  {\bibinfo  {journal} {Nature communications}\ }\textbf {\bibinfo {volume}
  {8}},\ \bibinfo {pages} {604} (\bibinfo {year} {2017})}\BibitemShut {NoStop}%
\bibitem [{\citenamefont {Peterson}\ \emph {et~al.}(2017)\citenamefont
  {Peterson}, \citenamefont {Lecocq}, \citenamefont {Cicak}, \citenamefont
  {Simmonds}, \citenamefont {Aumentado},\ and\ \citenamefont
  {Teufel}}]{peterson:2017}%
  \BibitemOpen
  \bibfield  {author} {\bibinfo {author} {\bibfnamefont {G.~A.}\ \bibnamefont
  {Peterson}}, \bibinfo {author} {\bibfnamefont {F.}~\bibnamefont {Lecocq}},
  \bibinfo {author} {\bibfnamefont {K.}~\bibnamefont {Cicak}}, \bibinfo
  {author} {\bibfnamefont {R.~W.}\ \bibnamefont {Simmonds}}, \bibinfo {author}
  {\bibfnamefont {J.}~\bibnamefont {Aumentado}}, \ and\ \bibinfo {author}
  {\bibfnamefont {J.~D.}\ \bibnamefont {Teufel}},\ }\href {\doibase
  10.1103/PhysRevX.7.031001} {\bibfield  {journal} {\bibinfo  {journal} {Phys.
  Rev. X}\ }\textbf {\bibinfo {volume} {7}},\ \bibinfo {pages} {031001}
  (\bibinfo {year} {2017})}\BibitemShut {NoStop}%
\bibitem [{\citenamefont {Barzanjeh}\ \emph {et~al.}(2017)\citenamefont
  {Barzanjeh}, \citenamefont {Wulf}, \citenamefont {Peruzzo}, \citenamefont
  {Kalaee}, \citenamefont {Dieterle}, \citenamefont {Painter},\ and\
  \citenamefont {Fink}}]{barzanjeh:2017}%
  \BibitemOpen
  \bibfield  {author} {\bibinfo {author} {\bibfnamefont {S.}~\bibnamefont
  {Barzanjeh}}, \bibinfo {author} {\bibfnamefont {M.}~\bibnamefont {Wulf}},
  \bibinfo {author} {\bibfnamefont {M.}~\bibnamefont {Peruzzo}}, \bibinfo
  {author} {\bibfnamefont {M.}~\bibnamefont {Kalaee}}, \bibinfo {author}
  {\bibfnamefont {P.~B.}\ \bibnamefont {Dieterle}}, \bibinfo {author}
  {\bibfnamefont {O.}~\bibnamefont {Painter}}, \ and\ \bibinfo {author}
  {\bibfnamefont {J.~M.}\ \bibnamefont {Fink}},\ }\href@noop {} {\bibfield
  {journal} {\bibinfo  {journal} {Nature Communications}\ }\textbf {\bibinfo
  {volume} {8}} (\bibinfo {year} {2017})}\BibitemShut {NoStop}%
\bibitem [{\citenamefont {Metelmann}\ and\ \citenamefont
  {T{\"u}reci}(2017)}]{metelmann:2017}%
  \BibitemOpen
  \bibfield  {author} {\bibinfo {author} {\bibfnamefont {A.}~\bibnamefont
  {Metelmann}}\ and\ \bibinfo {author} {\bibfnamefont {H.~E.}\ \bibnamefont
  {T{\"u}reci}},\ }\href@noop {} {\bibfield  {journal} {\bibinfo  {journal}
  {arXiv preprint arXiv:1703.04052}\ } (\bibinfo {year} {2017})}\BibitemShut
  {NoStop}%
\bibitem [{\citenamefont {Khorasani}(2017)}]{khorasani:2017}%
  \BibitemOpen
  \bibfield  {author} {\bibinfo {author} {\bibfnamefont {S.}~\bibnamefont
  {Khorasani}},\ }\href@noop {} {\bibfield  {journal} {\bibinfo  {journal}
  {IEEE Journal of Quantum Electronics}\ } (\bibinfo {year}
  {2017})}\BibitemShut {NoStop}%
\bibitem [{\citenamefont {Castellanos-Beltran}\ and\ \citenamefont
  {Lehnert}(2007)}]{castellanos:2007}%
  \BibitemOpen
  \bibfield  {author} {\bibinfo {author} {\bibfnamefont {M.~A.}\ \bibnamefont
  {Castellanos-Beltran}}\ and\ \bibinfo {author} {\bibfnamefont {K.~W.}\
  \bibnamefont {Lehnert}},\ }\href {\doibase 10.1063/1.2773988} {\bibfield
  {journal} {\bibinfo  {journal} {Applied Physics Letters}\ }\textbf {\bibinfo
  {volume} {91}},\ \bibinfo {pages} {083509} (\bibinfo {year}
  {2007})}\BibitemShut {NoStop}%
\bibitem [{\citenamefont {Mutus}\ \emph {et~al.}(2013)\citenamefont {Mutus},
  \citenamefont {White}, \citenamefont {Jeffrey}, \citenamefont {Sank},
  \citenamefont {Barends}, \citenamefont {Bochmann}, \citenamefont {Chen},
  \citenamefont {Chen}, \citenamefont {Chiaro}, \citenamefont {Dunsworth},
  \citenamefont {Kelly}, \citenamefont {Megrant}, \citenamefont {Neill},
  \citenamefont {O'Malley}, \citenamefont {Roushan}, \citenamefont
  {Vainsencher}, \citenamefont {Wenner}, \citenamefont {Siddiqi}, \citenamefont
  {Vijay}, \citenamefont {Cleland},\ and\ \citenamefont
  {Martinis}}]{mutus:2013}%
  \BibitemOpen
  \bibfield  {author} {\bibinfo {author} {\bibfnamefont {J.~Y.}\ \bibnamefont
  {Mutus}}, \bibinfo {author} {\bibfnamefont {T.~C.}\ \bibnamefont {White}},
  \bibinfo {author} {\bibfnamefont {E.}~\bibnamefont {Jeffrey}}, \bibinfo
  {author} {\bibfnamefont {D.}~\bibnamefont {Sank}}, \bibinfo {author}
  {\bibfnamefont {R.}~\bibnamefont {Barends}}, \bibinfo {author} {\bibfnamefont
  {J.}~\bibnamefont {Bochmann}}, \bibinfo {author} {\bibfnamefont
  {Y.}~\bibnamefont {Chen}}, \bibinfo {author} {\bibfnamefont {Z.}~\bibnamefont
  {Chen}}, \bibinfo {author} {\bibfnamefont {B.}~\bibnamefont {Chiaro}},
  \bibinfo {author} {\bibfnamefont {A.}~\bibnamefont {Dunsworth}}, \bibinfo
  {author} {\bibfnamefont {J.}~\bibnamefont {Kelly}}, \bibinfo {author}
  {\bibfnamefont {A.}~\bibnamefont {Megrant}}, \bibinfo {author} {\bibfnamefont
  {C.}~\bibnamefont {Neill}}, \bibinfo {author} {\bibfnamefont {P.~J.~J.}\
  \bibnamefont {O'Malley}}, \bibinfo {author} {\bibfnamefont {P.}~\bibnamefont
  {Roushan}}, \bibinfo {author} {\bibfnamefont {A.}~\bibnamefont
  {Vainsencher}}, \bibinfo {author} {\bibfnamefont {J.}~\bibnamefont {Wenner}},
  \bibinfo {author} {\bibfnamefont {I.}~\bibnamefont {Siddiqi}}, \bibinfo
  {author} {\bibfnamefont {R.}~\bibnamefont {Vijay}}, \bibinfo {author}
  {\bibfnamefont {A.~N.}\ \bibnamefont {Cleland}}, \ and\ \bibinfo {author}
  {\bibfnamefont {J.~M.}\ \bibnamefont {Martinis}},\ }\href@noop {} {\bibfield
  {journal} {\bibinfo  {journal} {Applied Physics Letters}\ }\textbf {\bibinfo
  {volume} {103}},\ \bibinfo {pages} {122602} (\bibinfo {year}
  {2013})}\BibitemShut {NoStop}%
\bibitem [{\citenamefont {Sauvageau}\ \emph {et~al.}(1995)\citenamefont
  {Sauvageau}, \citenamefont {Burroughs}, \citenamefont {Booi}, \citenamefont
  {Cromar}, \citenamefont {Benz},\ and\ \citenamefont {Koch}}]{sauvageau:1995}%
  \BibitemOpen
  \bibfield  {author} {\bibinfo {author} {\bibfnamefont {J.~E.}\ \bibnamefont
  {Sauvageau}}, \bibinfo {author} {\bibfnamefont {C.~J.}\ \bibnamefont
  {Burroughs}}, \bibinfo {author} {\bibfnamefont {P.~A.~A.}\ \bibnamefont
  {Booi}}, \bibinfo {author} {\bibfnamefont {M.~W.}\ \bibnamefont {Cromar}},
  \bibinfo {author} {\bibfnamefont {R.~P.}\ \bibnamefont {Benz}}, \ and\
  \bibinfo {author} {\bibfnamefont {J.~A.}\ \bibnamefont {Koch}},\ }\href@noop
  {} {\bibfield  {journal} {\bibinfo  {journal} {Applied Superconductivity,
  IEEE Transactions on}\ }\textbf {\bibinfo {volume} {5}},\ \bibinfo {pages}
  {2303} (\bibinfo {year} {1995})}\BibitemShut {NoStop}%
\bibitem [{\citenamefont {Mates}\ \emph {et~al.}(2008)\citenamefont {Mates},
  \citenamefont {Hilton}, \citenamefont {Irwin}, \citenamefont {Vale},\ and\
  \citenamefont {Lehnert}}]{mates:2008}%
  \BibitemOpen
  \bibfield  {author} {\bibinfo {author} {\bibfnamefont {J.~A.~B.}\
  \bibnamefont {Mates}}, \bibinfo {author} {\bibfnamefont {G.~C.}\ \bibnamefont
  {Hilton}}, \bibinfo {author} {\bibfnamefont {K.~D.}\ \bibnamefont {Irwin}},
  \bibinfo {author} {\bibfnamefont {L.~R.}\ \bibnamefont {Vale}}, \ and\
  \bibinfo {author} {\bibfnamefont {K.~W.}\ \bibnamefont {Lehnert}},\
  }\href@noop {} {\bibfield  {journal} {\bibinfo  {journal} {Applied Physics
  Letters}\ }\textbf {\bibinfo {volume} {92}},\ \bibinfo {pages} {023514}
  (\bibinfo {year} {2008})}\BibitemShut {NoStop}%
\bibitem [{\citenamefont {Tellegen}(1948)}]{tellegen:1948}%
  \BibitemOpen
  \bibfield  {author} {\bibinfo {author} {\bibfnamefont {B.~D.~H.}\
  \bibnamefont {Tellegen}},\ }\href@noop {} {\bibfield  {journal} {\bibinfo
  {journal} {Philips Res. Rep}\ }\textbf {\bibinfo {volume} {3}},\ \bibinfo
  {pages} {81} (\bibinfo {year} {1948})}\BibitemShut {NoStop}%
\bibitem [{\citenamefont {Rosenthal}\ \emph {et~al.}(2017)\citenamefont
  {Rosenthal}, \citenamefont {Chapman}, \citenamefont {Higginbotham},
  \citenamefont {Kerckhoff},\ and\ \citenamefont {Lehnert}}]{rosenthal:2017}%
  \BibitemOpen
  \bibfield  {author} {\bibinfo {author} {\bibfnamefont {E.~I.}\ \bibnamefont
  {Rosenthal}}, \bibinfo {author} {\bibfnamefont {B.~J.}\ \bibnamefont
  {Chapman}}, \bibinfo {author} {\bibfnamefont {A.~P.}\ \bibnamefont
  {Higginbotham}}, \bibinfo {author} {\bibfnamefont {J.}~\bibnamefont
  {Kerckhoff}}, \ and\ \bibinfo {author} {\bibfnamefont {K.~W.}\ \bibnamefont
  {Lehnert}},\ }\href {\doibase 10.1103/PhysRevLett.119.147703} {\bibfield
  {journal} {\bibinfo  {journal} {Phys. Rev. Lett.}\ }\textbf {\bibinfo
  {volume} {119}},\ \bibinfo {pages} {147703} (\bibinfo {year}
  {2017})}\BibitemShut {NoStop}%
\bibitem [{\citenamefont {Chapman}\ \emph {et~al.}(2016)\citenamefont
  {Chapman}, \citenamefont {Moores}, \citenamefont {Rosenthal}, \citenamefont
  {Kerckhoff},\ and\ \citenamefont {Lehnert}}]{chapman:2016}%
  \BibitemOpen
  \bibfield  {author} {\bibinfo {author} {\bibfnamefont {B.~J.}\ \bibnamefont
  {Chapman}}, \bibinfo {author} {\bibfnamefont {B.~A.}\ \bibnamefont {Moores}},
  \bibinfo {author} {\bibfnamefont {E.~I.}\ \bibnamefont {Rosenthal}}, \bibinfo
  {author} {\bibfnamefont {J.}~\bibnamefont {Kerckhoff}}, \ and\ \bibinfo
  {author} {\bibfnamefont {K.~W.}\ \bibnamefont {Lehnert}},\ }\href@noop {}
  {\bibfield  {journal} {\bibinfo  {journal} {Applied Physics Letters}\
  }\textbf {\bibinfo {volume} {108}},\ \bibinfo {pages} {222602} (\bibinfo
  {year} {2016})}\BibitemShut {NoStop}%
\bibitem [{\citenamefont {Chapman}\ \emph {et~al.}(2017)\citenamefont
  {Chapman}, \citenamefont {Rosenthal}, \citenamefont {Kerckhoff},
  \citenamefont {Vale}, \citenamefont {Hilton},\ and\ \citenamefont
  {Lehnert}}]{chapman:2017}%
  \BibitemOpen
  \bibfield  {author} {\bibinfo {author} {\bibfnamefont {B.~J.}\ \bibnamefont
  {Chapman}}, \bibinfo {author} {\bibfnamefont {E.~I.}\ \bibnamefont
  {Rosenthal}}, \bibinfo {author} {\bibfnamefont {J.}~\bibnamefont
  {Kerckhoff}}, \bibinfo {author} {\bibfnamefont {L.~R.}\ \bibnamefont {Vale}},
  \bibinfo {author} {\bibfnamefont {G.~C.}\ \bibnamefont {Hilton}}, \ and\
  \bibinfo {author} {\bibfnamefont {K.~W.}\ \bibnamefont {Lehnert}},\
  }\href@noop {} {\bibfield  {journal} {\bibinfo  {journal} {Applied Physics
  Letters}\ }\textbf {\bibinfo {volume} {110}},\ \bibinfo {pages} {162601}
  (\bibinfo {year} {2017})}\BibitemShut {NoStop}%
\bibitem [{\citenamefont {Likharev}(1986)}]{likharev:1986}%
  \BibitemOpen
  \bibfield  {author} {\bibinfo {author} {\bibfnamefont {K.~K.}\ \bibnamefont
  {Likharev}},\ }\href@noop {} {\emph {\bibinfo {title} {Dynamics of Josephson
  junctions and circuits}}}\ (\bibinfo  {publisher} {Gordon and Breach science
  publishers},\ \bibinfo {year} {1986})\BibitemShut {NoStop}%
\bibitem [{\citenamefont {Riste}\ \emph {et~al.}(2013)\citenamefont {Riste},
  \citenamefont {Dukalski}, \citenamefont {Watson}, \citenamefont {de~Lange},
  \citenamefont {Tiggelman}, \citenamefont {Blanter}, \citenamefont {Lehnert},
  \citenamefont {Schouten},\ and\ \citenamefont {DiCarlo}}]{riste:2013}%
  \BibitemOpen
  \bibfield  {author} {\bibinfo {author} {\bibfnamefont {D.}~\bibnamefont
  {Riste}}, \bibinfo {author} {\bibfnamefont {M.}~\bibnamefont {Dukalski}},
  \bibinfo {author} {\bibfnamefont {C.~A.}\ \bibnamefont {Watson}}, \bibinfo
  {author} {\bibfnamefont {G.}~\bibnamefont {de~Lange}}, \bibinfo {author}
  {\bibfnamefont {M.~J.}\ \bibnamefont {Tiggelman}}, \bibinfo {author}
  {\bibfnamefont {Y.~M.}\ \bibnamefont {Blanter}}, \bibinfo {author}
  {\bibfnamefont {K.~W.}\ \bibnamefont {Lehnert}}, \bibinfo {author}
  {\bibfnamefont {R.~N.}\ \bibnamefont {Schouten}}, \ and\ \bibinfo {author}
  {\bibfnamefont {L.}~\bibnamefont {DiCarlo}},\ }\href@noop {} {\bibfield
  {journal} {\bibinfo  {journal} {Nature}\ }\textbf {\bibinfo {volume} {502}},\
  \bibinfo {pages} {350} (\bibinfo {year} {2013})}\BibitemShut {NoStop}%
\bibitem [{\citenamefont {Hacohen-Gourgy}\ \emph {et~al.}(2016)\citenamefont
  {Hacohen-Gourgy}, \citenamefont {Martin}, \citenamefont {Flurin},
  \citenamefont {Ramasesh}, \citenamefont {Whaley},\ and\ \citenamefont
  {Siddiqi}}]{hacohen:2016}%
  \BibitemOpen
  \bibfield  {author} {\bibinfo {author} {\bibfnamefont {S.}~\bibnamefont
  {Hacohen-Gourgy}}, \bibinfo {author} {\bibfnamefont {L.~S.}\ \bibnamefont
  {Martin}}, \bibinfo {author} {\bibfnamefont {E.}~\bibnamefont {Flurin}},
  \bibinfo {author} {\bibfnamefont {V.~V.}\ \bibnamefont {Ramasesh}}, \bibinfo
  {author} {\bibfnamefont {K.~B.}\ \bibnamefont {Whaley}}, \ and\ \bibinfo
  {author} {\bibfnamefont {I.}~\bibnamefont {Siddiqi}},\ }\href@noop {}
  {\bibfield  {journal} {\bibinfo  {journal} {Nature}\ }\textbf {\bibinfo
  {volume} {538}},\ \bibinfo {pages} {491} (\bibinfo {year}
  {2016})}\BibitemShut {NoStop}%
\bibitem [{\citenamefont {Doerr}\ \emph {et~al.}(2011)\citenamefont {Doerr},
  \citenamefont {Dupuis},\ and\ \citenamefont {Zhang}}]{doerr:2011}%
  \BibitemOpen
  \bibfield  {author} {\bibinfo {author} {\bibfnamefont {C.~R.}\ \bibnamefont
  {Doerr}}, \bibinfo {author} {\bibfnamefont {N.}~\bibnamefont {Dupuis}}, \
  and\ \bibinfo {author} {\bibfnamefont {L.}~\bibnamefont {Zhang}},\ }\href
  {\doibase 10.1364/OL.36.004293} {\bibfield  {journal} {\bibinfo  {journal}
  {Opt. Lett.}\ }\textbf {\bibinfo {volume} {36}},\ \bibinfo {pages} {4293}
  (\bibinfo {year} {2011})}\BibitemShut {NoStop}%
\bibitem [{\citenamefont {Yang}\ \emph {et~al.}(2014)\citenamefont {Yang},
  \citenamefont {Galland}, \citenamefont {Liu}, \citenamefont {Tan},
  \citenamefont {Ding}, \citenamefont {Li}, \citenamefont {Bergman},
  \citenamefont {Baehr-Jones},\ and\ \citenamefont {Hochberg}}]{yang:2014}%
  \BibitemOpen
  \bibfield  {author} {\bibinfo {author} {\bibfnamefont {Y.}~\bibnamefont
  {Yang}}, \bibinfo {author} {\bibfnamefont {C.}~\bibnamefont {Galland}},
  \bibinfo {author} {\bibfnamefont {Y.}~\bibnamefont {Liu}}, \bibinfo {author}
  {\bibfnamefont {K.}~\bibnamefont {Tan}}, \bibinfo {author} {\bibfnamefont
  {R.}~\bibnamefont {Ding}}, \bibinfo {author} {\bibfnamefont {Q.}~\bibnamefont
  {Li}}, \bibinfo {author} {\bibfnamefont {K.}~\bibnamefont {Bergman}},
  \bibinfo {author} {\bibfnamefont {T.}~\bibnamefont {Baehr-Jones}}, \ and\
  \bibinfo {author} {\bibfnamefont {M.}~\bibnamefont {Hochberg}},\ }\href@noop
  {} {\bibfield  {journal} {\bibinfo  {journal} {Optics express}\ }\textbf
  {\bibinfo {volume} {22}},\ \bibinfo {pages} {17409} (\bibinfo {year}
  {2014})}\BibitemShut {NoStop}%
\bibitem [{\citenamefont {Lalumi\`ere}(2015)}]{kevinthesis}%
  \BibitemOpen
  \bibfield  {author} {\bibinfo {author} {\bibfnamefont {K.}~\bibnamefont
  {Lalumi\`ere}},\ }\emph {\bibinfo {title} {\'{E}lectrodynamique quantique en
  guide d'onde}},\ \href@noop {} {Ph.D. thesis},\ \bibinfo  {school}
  {Universit\'e de Sherbrooke} (\bibinfo {year} {2015})\BibitemShut {NoStop}%
\bibitem [{\citenamefont {Stan}\ \emph {et~al.}(2004)\citenamefont {Stan},
  \citenamefont {Field},\ and\ \citenamefont {Martinis}}]{martinis:2004}%
  \BibitemOpen
  \bibfield  {author} {\bibinfo {author} {\bibfnamefont {G.}~\bibnamefont
  {Stan}}, \bibinfo {author} {\bibfnamefont {S.~B.}\ \bibnamefont {Field}}, \
  and\ \bibinfo {author} {\bibfnamefont {J.~M.}\ \bibnamefont {Martinis}},\
  }\href {\doibase 10.1103/PhysRevLett.92.097003} {\bibfield  {journal}
  {\bibinfo  {journal} {Phys. Rev. Lett.}\ }\textbf {\bibinfo {volume} {92}},\
  \bibinfo {pages} {097003} (\bibinfo {year} {2004})}\BibitemShut {NoStop}%
\bibitem [{\citenamefont {Van~Duzer}\ and\ \citenamefont
  {Turner}(1981)}]{vanduzer:1981}%
  \BibitemOpen
  \bibfield  {author} {\bibinfo {author} {\bibfnamefont {T.}~\bibnamefont
  {Van~Duzer}}\ and\ \bibinfo {author} {\bibfnamefont {C.~W.}\ \bibnamefont
  {Turner}},\ }\href@noop {} {\emph {\bibinfo {title} {Principles of
  superconductive devices and circuits}}},\ \bibinfo {edition} {2nd}\ ed.\
  (\bibinfo  {publisher} {Prentice Hall},\ \bibinfo {year} {1981})\BibitemShut
  {NoStop}%
\bibitem [{\citenamefont {Ashcroft}\ and\ \citenamefont
  {Mermin}(1976)}]{ashcroft:1976}%
  \BibitemOpen
  \bibfield  {author} {\bibinfo {author} {\bibfnamefont {N.~W.}\ \bibnamefont
  {Ashcroft}}\ and\ \bibinfo {author} {\bibfnamefont {N.~D.}\ \bibnamefont
  {Mermin}},\ }\href@noop {} {\emph {\bibinfo {title} {Solid state physics}}}\
  (\bibinfo  {publisher} {Holt, Rinehart and Winston},\ \bibinfo {year}
  {1976})\BibitemShut {NoStop}%
\bibitem [{\citenamefont {Pothier}\ \emph {et~al.}(1994)\citenamefont
  {Pothier}, \citenamefont {Gu\'eron}, \citenamefont {Esteve},\ and\
  \citenamefont {Devoret}}]{pothier:1994}%
  \BibitemOpen
  \bibfield  {author} {\bibinfo {author} {\bibfnamefont {H.}~\bibnamefont
  {Pothier}}, \bibinfo {author} {\bibfnamefont {S.}~\bibnamefont {Gu\'eron}},
  \bibinfo {author} {\bibfnamefont {D.}~\bibnamefont {Esteve}}, \ and\ \bibinfo
  {author} {\bibfnamefont {M.~H.}\ \bibnamefont {Devoret}},\ }\href {\doibase
  10.1103/PhysRevLett.73.2488} {\bibfield  {journal} {\bibinfo  {journal}
  {Phys. Rev. Lett.}\ }\textbf {\bibinfo {volume} {73}},\ \bibinfo {pages}
  {2488} (\bibinfo {year} {1994})}\BibitemShut {NoStop}%
\bibitem [{\citenamefont {Dubos}\ \emph {et~al.}(2001)\citenamefont {Dubos},
  \citenamefont {Courtois}, \citenamefont {Pannetier}, \citenamefont {Wilhelm},
  \citenamefont {Zaikin},\ and\ \citenamefont {Sch\"on}}]{dubos:2001}%
  \BibitemOpen
  \bibfield  {author} {\bibinfo {author} {\bibfnamefont {P.}~\bibnamefont
  {Dubos}}, \bibinfo {author} {\bibfnamefont {H.}~\bibnamefont {Courtois}},
  \bibinfo {author} {\bibfnamefont {B.}~\bibnamefont {Pannetier}}, \bibinfo
  {author} {\bibfnamefont {F.~K.}\ \bibnamefont {Wilhelm}}, \bibinfo {author}
  {\bibfnamefont {A.~D.}\ \bibnamefont {Zaikin}}, \ and\ \bibinfo {author}
  {\bibfnamefont {G.}~\bibnamefont {Sch\"on}},\ }\href {\doibase
  10.1103/PhysRevB.63.064502} {\bibfield  {journal} {\bibinfo  {journal} {Phys.
  Rev. B}\ }\textbf {\bibinfo {volume} {63}},\ \bibinfo {pages} {064502}
  (\bibinfo {year} {2001})}\BibitemShut {NoStop}%
\bibitem [{\citenamefont {Mohanty}\ and\ \citenamefont
  {Webb}(2003)}]{mohanty:2003}%
  \BibitemOpen
  \bibfield  {author} {\bibinfo {author} {\bibfnamefont {P.}~\bibnamefont
  {Mohanty}}\ and\ \bibinfo {author} {\bibfnamefont {R.~A.}\ \bibnamefont
  {Webb}},\ }\href {\doibase 10.1103/PhysRevLett.91.066604} {\bibfield
  {journal} {\bibinfo  {journal} {Phys. Rev. Lett.}\ }\textbf {\bibinfo
  {volume} {91}},\ \bibinfo {pages} {066604} (\bibinfo {year}
  {2003})}\BibitemShut {NoStop}%
\bibitem [{\citenamefont {Scully}\ and\ \citenamefont
  {Zubairy}(1997)}]{scully:1997}%
  \BibitemOpen
  \bibfield  {author} {\bibinfo {author} {\bibfnamefont {M.~O.}\ \bibnamefont
  {Scully}}\ and\ \bibinfo {author} {\bibfnamefont {M.~S.}\ \bibnamefont
  {Zubairy}},\ }\href@noop {} {\emph {\bibinfo {title} {Quantum optics}}}\
  (\bibinfo  {publisher} {Cambridge University Press},\ \bibinfo {year}
  {1997})\ pp.\ \bibinfo {pages} {101--104}\BibitemShut {NoStop}%
\bibitem [{\citenamefont {Johnson}(1928)}]{johnson:1928}%
  \BibitemOpen
  \bibfield  {author} {\bibinfo {author} {\bibfnamefont {J.~B.}\ \bibnamefont
  {Johnson}},\ }\href {\doibase 10.1103/PhysRev.32.97} {\bibfield  {journal}
  {\bibinfo  {journal} {Phys. Rev.}\ }\textbf {\bibinfo {volume} {32}},\
  \bibinfo {pages} {97} (\bibinfo {year} {1928})}\BibitemShut {NoStop}%
\bibitem [{\citenamefont {Nyquist}(1928)}]{nyquist:1928}%
  \BibitemOpen
  \bibfield  {author} {\bibinfo {author} {\bibfnamefont {H.}~\bibnamefont
  {Nyquist}},\ }\href {\doibase 10.1103/PhysRev.32.110} {\bibfield  {journal}
  {\bibinfo  {journal} {Phys. Rev.}\ }\textbf {\bibinfo {volume} {32}},\
  \bibinfo {pages} {110} (\bibinfo {year} {1928})}\BibitemShut {NoStop}%
\end{thebibliography}
%

\end{document}